\documentclass[twocolumn, numberedappendix,appendixfloats]{openjournal}
\usepackage{lineno}
\usepackage[frozencache]{minted}
\usepackage{listings}
\usepackage{booktabs}
\lstdefinestyle{yaml}{
     basicstyle=\ttfamily\color{black}\scriptsize,
     rulecolor=\color{black},
     keywordstyle=\color{blue}\bfseries,
     comment=[l]{\#},
     commentstyle=\itshape\color{black},
     showstringspaces=false
}
\usepackage{lipsum}
\newcommand{\pyimcom}{\texttt{pyimcom}}
\newcommand{\slimfarmer}{\texttt{slimfarmer}}
\newcommand{\Dcsim}{\texttt{DC25 sim}}
\newcommand{\ou}{\texttt{OpenUniverse2024}}

\usepackage{enumitem}

\usepackage{xcolor}
\usepackage{textgreek}
\usepackage[utf8]{inputenc}
\usepackage[english]{babel}

\usepackage{hyperref}
\hypersetup{
    unicode, 
    colorlinks=true,
    linkcolor=linkcolor,
    citecolor=linkcolor,
    filecolor=linkcolor,
    urlcolor=linkcolor,
}
\usepackage{color,colortbl}
\definecolor{linkcolor}{rgb}{0.0,0.3,0.5}
\usepackage{tensind}
\tensordelimiter{?}
\DeclareGraphicsExtensions{.bmp,.png,.jpg,.pdf}
\usepackage{verbatim}
\usepackage[normalem]{ulem}
\usepackage{orcidlink}
\usepackage{soul}

\graphicspath{ {./figs/} }
\makeatletter
\renewcommand{\andname}{for the}
\makeatother
\begin{document}
\title{Robust Photometry for Roman High-Latitude Imaging Survey Cosmology Using Roman and Rubin Imaging} 
\shorttitle{Roman HLIS-PIT photometry}
\shortauthors{To et al.}

\author{Chun-Hao To$^{*}$\orcidlink{0000-0001-7836-2261}}
\affiliation{Astronomy and Astrophysics Department, The University of Chicago}
\affiliation{Kavli Institute for Cosmological Physics, University of Chicago, Chicago, IL 60637, USA}
\affiliation{NSF-Simons AI Institute for the Sky (SkAI),172 E. Chestnut St., Chicago, IL 60611, USA}
\email{$^*$email: chto@uchicago.edu}
\author{Tae-Hyeon Shin\orcidlink{0000-0002-6389-5409}}
\affiliation{McWilliams Center for Cosmology and Astrophysics, Department of Physics, Carnegie Mellon University, Pittsburgh, PA 15213, USA}

\author{Chihway Chang\orcidlink{0000-0002-7887-0896}}
\affiliation{Astronomy and Astrophysics Department, The University of Chicago}
\affiliation{Kavli Institute for Cosmological Physics, University of Chicago, Chicago, IL 60637, USA}
\affiliation{NSF-Simons AI Institute for the Sky (SkAI),172 E. Chestnut St., Chicago, IL 60611, USA}

\author{Christopher Hirata}
\affiliation{Center for Cosmology and AstroParticle Physics (CCAPP), The Ohio State University, 191 West Woodruff Ave, Columbus, OH 43210, USA}
\affiliation{Department of Astronomy, The Ohio State University, 140 West 18th Avenue, Columbus, OH 43210, USA}
\affiliation{Department of Physics, The Ohio State University, 191 West Woodruff Ave, Columbus, OH 43210, USA
}

\author{Brett H. Andrews\orcidlink{0000-0001-8085-5890}}
\affiliation{Department of Physics and Astronomy, University of Pittsburgh, Pittsburgh, PA 15260, USA}
\affiliation{Pittsburgh Particle Physics, Astrophysics, and Cosmology Center (PITT PACC), University of Pittsburgh, Pittsburgh, PA 15260, USA}
\author{Ami Choi \orcidlink{0000-0002-5636-233X}}
\affiliation{NASA Goddard Space Flight Center, Greenbelt, MD 20771, USA}

\author{Boyan Yin}
\affiliation{Department of Physics, Duke University, Durham, NC 27708, USA}
\author{Rachel Mandelbaum\orcidlink{0000-0003-2271-1527}}
\affiliation{McWilliams Center for Cosmology and Astrophysics, Department of Physics, Carnegie Mellon University, Pittsburgh, PA 15213, USA}

\author{Federico Berlfein\orcidlink{0009-0004-4113-9938}}
\affiliation{McWilliams Center for Cosmology and Astrophysics, Department of Physics, Carnegie Mellon University, Pittsburgh, PA 15213, USA}

\author{Kaili Cao\orcidlink{0000-0002-1699-6944}
}
\affiliation{Center for Cosmology and AstroParticle Physics (CCAPP), The Ohio State University, 191 West Woodruff Ave, Columbus, OH 43210, USA}
\affiliation{Department of Physics, The Ohio State University, 191 West Woodruff Ave, Columbus, OH 43210, USA
}
\author{James Chiang\orcidlink{0000-0001-5738-8956}}
\affiliation{SLAC National Accelerator Laboratory, 2575 Sand Hill Road, Menlo Park, CA, 94025, USA}
\affiliation{Kavli Institute for Particle Astrophysics and Cosmology, Stanford University, Stanford CA 94305, USA}

\author{Nihar Dalal\orcidlink{0000-0003-2177-9407}}
\affiliation{Center for Cosmology and AstroParticle Physics (CCAPP), The Ohio State University, 191 West Woodruff Ave, Columbus, OH 43210, USA}
\affiliation{Department of Physics, The Ohio State University, 191 West Woodruff Ave, Columbus, OH 43210, USA
}

\author{Yuedong Fang\orcidlink{0000-0002-0334-6950}}
\affiliation{Department of Physics, Duke University, Durham, NC 27708, USA}

\author{Axel Guinot\orcidlink{0000-0002-5068-7918}}
\affiliation{McWilliams Center for Cosmology and Astrophysics, Department of Physics, Carnegie Mellon University, Pittsburgh, PA 15213, USA}

\author{Michael Gabe}
\affiliation{Center for Cosmology and AstroParticle Physics (CCAPP), The Ohio State University, 191 West Woodruff Ave, Columbus, OH 43210, USA}

\author{Andrew Hearin\orcidlink{0000-0003-2219-6852}
}
\affiliation{HEP Division, Argonne National Laboratory, 9700 South Cass Avenue, Lemont, IL 60439, USA}

\author{Katherine Laliotis\orcidlink{0000-0002-6111-6061}}
\affiliation{Center for Cosmology and AstroParticle Physics (CCAPP), The Ohio State University, 191 West Woodruff Ave, Columbus, OH 43210, USA}
\affiliation{Department of Physics, The Ohio State University, 191 West Woodruff Ave, Columbus, OH 43210, USA
}

\author{Emily Macbeth\orcidlink{0009-0002-7190-9775}}
\affiliation{Steward Observatory \& Department of Astronomy, University of Arizona, 933 North Cherry Avenue, Tucson, AZ 85721, USA}
\affiliation{Center for Cosmology and AstroParticle Physics (CCAPP), The Ohio State University, 191 West Woodruff Ave, Columbus, OH 43210, USA}
\affiliation{Department of Physics, The Ohio State University, 191 West Woodruff Ave, Columbus, OH 43210, USA
}
\affiliation{Department of Astronomy, The Ohio State University, 140 West 18th Avenue, Columbus, OH 43210, USA}

\author{Sidney Mau\orcidlink{0000-0003-3519-4004}}
\affiliation{Department of Physics, Duke University, Durham, NC 27708, USA}

\author{Edward F. Schlafly\orcidlink{0000-0002-3569-7421}}
\affiliation{Space Telescope Science Institute}
\author{M.~A.~Troxel\orcidlink{0000-0002-5622-5212}}
\affiliation{Department of Physics, Duke University, Durham, NC 27708, USA}
\author{Roman HLIS Cosmology Project Infrastructure Team}
\noaffiliation

\begin{abstract}
Accurate and precise photometry is essential for Roman cosmology because it affects photometric-redshift performance, galaxy and cluster sample selection, tomographic binning, and redshift-distribution characterization. Achieving robust photometry is challenging because Roman’s depth leads to substantial source blending, particularly in joint Roman--Rubin analyses, where the two surveys have different spatial resolutions. In this work, we develop and validate \texttt{slimfarmer}, a model-fitting photometry pipeline designed for Roman High-Latitude Imaging Survey cosmological analyses. Building on profile-fitting photometry package \texttt{The Farmer}, \texttt{slimfarmer} introduces treatments of the correlated noise present in Roman coadded images and of astronomical shot noise, together with model-fitting configurations tuned for Roman imaging. We validate the pipeline using Roman coadded simulated images based on \texttt{OpenUniverse2024} and perform forced photometry on matched Rubin image simulations. We find that \texttt{slimfarmer} recovers galaxy colors with the mode of the color residuals within $20$ millimag for Roman-based colors and within $70$ millimag for Rubin-based colors. Properly accounting for correlated noise is essential for uncertainty quantification, as photometric uncertainties are otherwise underestimated by up to a factor of $\sim 3$. 
We also find that applying multi-object fitting consistently to both Roman and Rubin imaging substantially mitigates blending-induced color systematics. In crowded regions, single-object fitting produces environment-dependent color biases of up to $\sim 0.5$ mag, leading to correspondingly stronger environment-dependent degradation in photometric-redshift performance.
Together, these results establish \texttt{slimfarmer} as a prototype photometric pipeline for Roman High-Latitude Imaging Survey cosmology applications and identify the correlated-noise corrections, astronomical shot-noise treatment, and joint Roman--Rubin multi-object fitting required for robust photometric-redshift characterization.
\end{abstract}


\maketitle

\section{Introduction}
\label{sec:intro}
The Wide Field Instrument on the Nancy Grace Roman Space Telescope (Roman) will provide high-resolution near-infrared imaging over large areas ($\sim5000\,\rm{deg}^2$), enabling transformative advances in studies of dark matter, dark energy, galaxy evolution, and many other areas of astrophysics \citep{ 2015arXiv150303757S,2019arXiv190205569A,timeallocation}. For cosmology, Roman will enable precision measurements of weak gravitational lensing, galaxy clustering, galaxy clusters, and their cross-correlations \citep{tim2021,Fishercpip,Junzo}. Translating these high-quality images into scientific discoveries, however, will depend critically on the quality of the derived catalogs. Galaxy photometry\footnote{Astronomical photometry generally refers to measuring the fluxes of astronomical objects, including both point sources and extended sources. In this paper, we focus specifically on photometry of galaxies.} is a central catalog-level data product for Roman cosmology analyses with the High-Latitude Imaging Survey (HLIS). It enters galaxy sample selection, cluster sample definitions, tomographic redshift binning, and the calibration of redshift distributions. Robust Roman HLIS cosmology, therefore, requires photometric measurements with well-characterized biases, uncertainties, and dependencies on observing conditions and galaxy environment. 

Astronomical photometry has a long history, and many methods have been developed to measure galaxy fluxes and colors from imaging data. These methods can be broadly divided into two categories \citep[e.g.,][for a discussion]{Farmer}. The first category consists of aperture-based methods, which measure the light enclosed within a fixed or adaptive region \citep[e.g.,][]{1980ApJS...43..305K, sextractor, 2008A&A...482.1053K}. These methods are often simple and computationally efficient, but they require aperture corrections to estimate total fluxes. In addition, color measurements can be biased when the point spread function (PSF), depth, or pixel scale differs across bands, a challenge that is especially important for joint analyses of Roman and other ground-based experiments, such as the Rubin Observatory's Legacy Survey of Space and Time (LSST, \citealt{lsst}). The second category consists of model-fitting approaches, in which galaxies are described by surface-brightness models, either parametric or nonparametric, convolved with the appropriate PSF \citep{galfit, galfit2, imfit, tractor, scarlet, sextractorplusplus, scarlet2}. By explicitly accounting for the PSF and the light profiles of neighboring sources, these methods can estimate total fluxes more directly and model blended galaxies simultaneously, making them particularly useful for deep imaging data where source blending is significant. One widely used implementation of this general approach is \texttt{The Farmer} \citep{Farmer}, an automated profile-fitting photometry package built around \texttt{The Tractor} \citep{tractor}. \texttt{The Farmer} has been applied to deep multiwavelength extragalactic catalogs  \citep{cosmos20,2023ApJS..269...46L,Euclid_prelaunch}, while related \texttt{The Tractor}-based photometry approaches have been used in SPHEREx simulations and data-processing development \citep[e.g.,][]{2024ApJ...972...68F,2025arXiv251115823A,2026ApJ..1000...56H}.

Roman images also present several additional challenges for accurate photometry compared to ground-based telescopes. First, the readout electronics of Roman’s Teledyne H4RG detectors can introduce correlated noise across pixels \citep{2012SPIE.8453E..1FR, noisecorrelation}\footnote{The total noise correlation includes contributions from both detector-level effects and the HLIS-PIT image-coaddition procedure. In appendix~\ref{app:noise_correlation}, we quantify the relative contributions of these two sources.}. If these correlations are ignored, the flux uncertainties can be underestimated, leading to overconfident color measurements and potentially biased downstream photometric-redshift inference. Second, unlike in many ground-based observations, where sky background often dominates, the astronomical shot noise from the galaxies themselves contributes non-negligibly to the total noise budget for Roman weak-lensing galaxy samples and must be treated carefully in model-fitting photometry. Third, the planned Roman High-Latitude Imaging Survey (HLIS, \citealt{timeallocation}) is expected to produce galaxy samples with exceptionally high number densities ($\sim 40/\rm{arcmin}^2$), making source blending an important challenge. Roman's high angular resolution will help mitigate blending in the near-infrared imaging, but this issue becomes more complicated in joint analyses with ground-based data, where the broader PSF leads to substantially more blended sources \citep{2021JCAP...07..043S,2018PASJ...70S...5B,2025rubn.rept...33A}. Blending can bias not only total flux measurements but also galaxy colors, especially when neighboring sources are treated differently across bands or surveys. As a result, blending can introduce environment-dependent color errors that propagate into sample selection, photometric redshift estimation, and downstream cosmological analyses. 

To address these challenges, we develop and validate \slimfarmer{}, a model-fitting photometry pipeline designed for Roman cosmological analyses. We place particular emphasis on joint Roman--Rubin photometry, which we show below is essential for Roman HLIS cosmology. \texttt{Slimfarmer} follows the general philosophy of \texttt{The Farmer} \citep{Farmer}: sources are detected in Roman coadded images, neighboring galaxies are grouped using dilated segmentation maps, and multi-band photometry is measured with \texttt{The Tractor} \citep{tractor} through simultaneous fitting of all sources within each group. This multi-object fitting approach allows blended galaxies to be modeled consistently, which is essential for avoiding environment-dependent color biases. We extend this framework in several ways that are important for Roman. First, we propagate correlated noise in the Roman coadds into the photometric uncertainties using noise realizations processed through the same image-processing pipeline as the science images. Second, we include a model-based treatment of astronomical shot noise in the inverse-variance weight maps, avoiding the biases that can arise when the noisy science image itself is used to estimate the source Poisson contribution (see appendix~\ref{app:shot_noise}). Third, we use the Roman-derived galaxy models to perform forced photometry on matched Rubin images, enabling consistent Roman--Rubin color measurements despite differences in PSFs, pixel scales, depths, and blending properties between the two surveys. As demonstrated by \citet{Spitzerforce}, forced photometry using higher-resolution imaging as a prior can provide more accurate measurements of faint sources than catalog-level matching alone.

The goal of \slimfarmer{} is to deliver photometric fluxes, colors, and uncertainty estimates that are reliable for Roman HLIS cosmology applications in which photometry enters through galaxy selection and photometric-redshift characterization. These include galaxy clustering analyses, galaxy cluster identification, and source-galaxy redshift calibration, as discussed above. In this work, we test this framework on Roman and Rubin image simulations: we validate the recovered photometry against the simulation truth catalog, quantify its dependence on galaxy environment, and propagate different photometric choices, including single-object fitting and the neglect of correlated noise, into both ensemble and individual photometric-redshift estimates. 
For weak-lensing cosmology applications, however, the present work does not propagate these photometric measurements through a \texttt{METADETECTION}-based framework \citep{2020ApJ...902..138S}, which will be required for end-to-end weak-lensing validation and is left for future work. Accordingly, statements about the role of multi-object fitting in this paper refer to photometric fluxes, colors, and photometric-redshift characterization. They should not be interpreted as conclusions about the optimal treatment of blending for Roman weak-lensing cosmology, for which the impact of different blending treatments on shear measurement must also be considered \citep[e.g.,][]{2023OJAp....6E..17S}. Finally, although this validation is performed on products generated by the Roman High-Latitude Imaging Survey Project Infrastructure Team (HLIS-PIT) processing pipeline, the methods developed here are not specific to those products and should also be applicable to imaging products produced by the Roman Science Operations Center.

This paper is organized as follows. Section~\ref{sec:sim} describes the Roman and Rubin image simulations used in this work. Section~\ref{sec:photomeasurement} presents the \slimfarmer{} detection, model-fitting, forced-photometry, and uncertainty-estimation framework. Section~\ref{sec:absolutephotometyandcolor} validates the measured total fluxes and colors against the simulation truth catalog and examines how alternative photometric choices, including single-object fitting and ignoring correlated noise, affect flux and color recovery. Section~\ref{sec:totalredshift} studies how these photometric choices propagate into photometric-redshift performance, including individual photo-$z$ scatter and outlier rate, redshift-distribution characterization, and tomographic binning. We summarize our conclusions in Section~\ref{sec:conclude}. Throughout the paper, the magnitudes are expressed in the AB system \citep{1983ApJ...266..713O}.

\section{Simulations}
\label{sec:sim}
\subsection{Roman image simulation}
In this work, we use the \Dcsim{}\footnote{Note that this is independent of the simulation produced by the Roman Science Operations Center.}, a Roman coadded-image simulation based on galaxy truth images from \ou{} \citep{OU24}. The simulation is divided into four equal-area mosaics, each spanning $1.23\ \deg^2$. In this paper, we analyze one of these mosaics, which includes realizations in the Y106, J129, H158, F184, and K213 bands.

The images are processed using a substantial subset of the Roman High-Latitude Imaging Survey Project Infrastructure Team (HLIS-PIT) pipeline; further details will be provided in \cite{DC25simrelease}. We briefly summarize the processing steps below.
\begin{enumerate}
    \item \textbf{Level 1 uncalibrated detector data}: 
    The simulated detector images include the following non-ideal physical effects:
    pupil variation, geometric distortion, optical aberrations, charge diffusion, cosmic rays, bias and dark currents, reset noise, sky background, flat-field variations, inter-pixel capacitance, gain, classical non-linearity, saturation, uncorrelated and correlated read noise, and up-the-ramp sampling. These effects are modeled using inputs from laboratory tests and the following codes: \texttt{romanimpreprocess v$0.1$}, \texttt{roman\_datamodels v$0.28.1$}, and \texttt{romanisim v$0.11.2$}.

    \item \textbf{Level 2 calibrated detector rate images}: 
    The Level 1 images are calibrated through reference-pixel correction, bias correction, classical linearity correction, dark-current subtraction, inter-pixel capacitance correction, ramp fitting \citep{2022romanrept..394C}, jump detection \citep{cosmicray}, flat-field correction, and sky subtraction. These steps are performed using \texttt{romanimpreprocess v$0.1$} and \texttt{stcal v$1.15.2$}.

    \item \textbf{Level 2.1 calibrated detector rate images}: 
    We apply two additional processing steps: removal of correlated read noise \citep[ImDestripe][]{imdestripe} and masking of outlier pixels whose values differ significantly from overlapping observations. These steps are performed using \texttt{roman\_hlis\_l2\_driver v$0.1.1$}.

    \item \textbf{Level 3 coadded images}: 
    The calibrated images are coadded using \texttt{pyimcom} \citep{Rowe,imcom, Kailiimcom}, a linear-algebra-based coaddition algorithm that combines multiple undersampled images into a single oversampled output mosaic with a specified target point spread function.
\end{enumerate}

The final processed data consist of galaxy images in the Y106, J129, H158, F184, and K213 bands covering $1.23\ \rm{deg}^2$. This image is further divided into $1600$ $1.66'\times 1.66'$ blocks. The coadds have a Gaussian PSF \citep{2026ApJ...998..304C} with full width at half maximum (FWHM) of $0.24''$ in Y106, J129, H158, and F184, and $0.27''$ in K213. The output pixel scale is $0.049''$. 

In addition to the galaxy images, denoted by $d$, we also generate several associated noise fields. These include four realizations of a background-only noise field, $N_{\rm bg}$, which contains background and instrumental noise; four realizations of a de-biasing field\footnote{This field is developed to remove biases caused by astronomical-object shot noise. Hence, it is called de-biasing field.}, $N_{\rm db}$, which contains background, instrumental, and astronomical-object shot noise; and one white-noise realization drawn from a unit normal distribution. These fields are passed through the same Level 2--3 processing steps as the science images, allowing us to study the noise correlations introduced by the calibration and coaddition procedures.

This simulation currently does not include several known Roman WFS detector effects, such as the Brighter-fatter effect \citep{bfe1, bfe2, bfe3}, persistence \citep{persistence}, optical ghosts, vertical trailing pixel effect \citep{bfe3}, and count-rate-dependent nonlinearity \citep{2021RNAAS...5...66D}. While the brighter-fatter effect can cause additional correlation between fluxes in neighboring pixels, we expect the impact on the photometry of the majority of weak lensing samples to  be negligible given the pixel area change by $<0.01\%$.

\subsection{Rubin image simulation}
The same set of galaxies in \Dcsim{} were also simulated in the Rubin ugrizy bands using the LSST DESC \texttt{imSim} framework. These images correspond to the first five years of the simulated LSST Wide-Fast-Deep observing sequence, as defined by OpSim v3.2, available in August 2023 \citep{OU24}. Starting from the same truth images used for the Roman simulations, \texttt{imSim} renders individual Rubin exposures while accounting for the observing conditions, wavelength-dependent system throughput, optical effects, and detector effects. The resulting raw exposures are processed with the LSST Science Pipelines \citep{10.71929/rubin/2570545,2018PASJ...70S...5B,2019ASPC..523..521B}, with data access handled through the Rubin Data Butler \citep{2022SPIE12189E..11J}. The calibrated exposures are resampled and combined into coadded images with a pixel scale of $0.2''$, accompanied by coadded PSF models, WCS solutions, variance maps, and masks.

\section{Photometry measurement}
\label{sec:photomeasurement}
This section describes our photometry measurement framework. Our pipeline closely follows 
\texttt{The Farmer}\footnote{\url{https://github.com/astroweaver/the_farmer}} \citep{Farmer} v1.0.0-beta: sources are detected using \texttt{SEP} v1.4.1\footnote{\url{https://github.com/sep-developers/sep}}
\citep{sextractor, sep}, and photometry is measured through multi-object model fitting 
using \texttt{The Tractor} v10.4.dev11 \citep{tractor}. 

\subsection{Detections}
We perform source detection on a combined detection image constructed from the mean of the 
Y106, J129, H158 coadded images, after scaling them to the same AB magnitude units. In addition to detecting 
sources, \texttt{SEP} produces a segmentation map that assigns each image pixel to a detected 
source. We use this segmentation map to define groups of neighboring galaxies that should be 
fit simultaneously. Specifically, we dilate each source's segmentation region by $0.2''$ and 
group galaxies whose dilated regions overlap. Among galaxies with signal-to-noise ratio greater than $18$, $42\%$ have dilated segmentation regions that overlap with those of at least one neighboring galaxy. The configuration of \texttt{SEP} used in this paper is detailed in appendix~\ref{app:detection}.

\subsection{Photometric measurements}
Given these source groups, we perform joint multi-band model fitting across the Y106, J129, H158, F184, and K213 bands using \texttt{The Tractor}. The inverse-variance weight maps are constructed following appendix~\ref{app:weight}, including a model-dependent treatment of astronomical shot noise. Because the astronomical shot-noise contribution depends on the galaxy model, it is updated at each fitting iteration using the current model prediction. We quantify the impact of this choice on the resulting photometric measurements in appendix~\ref{app:shot_noise}. For a given band $b$ 
and a group of galaxies $g$, we denote the science images, pixelized PSF, inverse-variance 
weight map, and model galaxy profiles with parameter $\theta$ as $d_{b}$, $\rm{PSF}_b$, $w_b$, and $m_{i\in g,b} (\theta)$, 
respectively. The log-likelihood is then given by
\begin{eqnarray}
\label{eq:mainlikelihood}
    -2\log \mathcal{L} &=& \sum_{b \in \rm{bands}} w_b\left(d_{b}-\sum_{i\in g}m_{i,b}(\theta)\ast\rm{PSF}_b \right)^2,
\end{eqnarray}
where bands are Y106, J129, H158, F184, K213, and $\ast$ denotes convolution, which is performed using the method described in \cite{2020arXiv201215797L}. 
Following \texttt{the Farmer}, we consider the following models for $m_{i,b}(\theta)$: 
\begin{enumerate}
    \item \textbf{Point Source} (PS): A delta-function profile specified by its flux and position in the celestial coordinate.

    \item \textbf{SimpleGalaxy} (SG): A circular exponential profile with a fixed half-light radius of $0.45''$. This model is specified only by its flux and position.

    \item \textbf{ExpGalaxy} (Exp): An exponential (lux) profile in the SDSS convention \citep{HoggandLang}, typically used to describe disk-like galaxies. This model is specified by its flux, position, effective radius, axis ratio, and position angle. To reduce the computational cost, the profile is approximated by a six-component Gaussian mixture model \citep{HoggandLang}.

    \item \textbf{DevGalaxy} (Dev): A de Vaucouleurs (luv) profile in the SDSS convention \citep{HoggandLang}, typically used to describe elliptical galaxies. This model is specified by the same set of parameters as ExpGalaxy: flux, position, effective radius, axis ratio, and position angle. Similarly, the profile is approximated by an eight-component Gaussian mixture model \citep{HoggandLang}.

    \item \textbf{FixedCompositeGalaxy} (Comp): A linear combination of ExpGalaxy and DevGalaxy components that share the same center. Each component has its own effective radius, axis ratio, and position angle. The total flux is specified, along with an additional parameter, $\mathrm{fracDev}$, which gives the fraction of the total flux assigned to the DevGalaxy component.
\end{enumerate}
Note that the position of the modeled galaxy is defined in the celestial coordinate, which is then transformed to the image coordinate using the World Coordinate System (WCS) associated to the coadded image. 

\subsubsection{Model selections}
Given the five possible models, an important task is to decide which is most suitable for a given galaxy. Following the procedure implemented in \texttt{The Farmer} \citep[see their figure~4 for a flowchart]{Farmer}, we perform model selection using a $\chi^2$-based decision tree. For each source, we fit the five candidate models sequentially using the likelihood defined in equation \ref{eq:mainlikelihood} and record the reduced chi-square $\chi^2_{\nu,M}$,
\begin{equation}
\label{eq:model_selection}
    \chi^2_{\nu,M} = \frac{1}{N_{\rm pix}-N_{\rm par}}\sum_{i\in S, b\in \rm{bands}}w_{i,b}\left(d_{i,b}-m_{i,b}^{(M)}(\hat{\theta})\ast\rm{PSF}_b\right)^2,
\end{equation}
where $M\in\{\rm{PS, SG, Exp, Dev, Comp}\}$ denotes the candidate model, $\hat{\theta}$ are the best-fit parameters at that stage, $S$ is the source's segmentation footprint, $N_{\rm pix}$ is the number of pixels in the source's segmentation footprint, and $N_{\rm par}$ is the number of free parameters in the fit for the specific source. Within a group, all sources are fit jointly at every stage, so that flux from neighbors is properly accounted for. Sources that are decided at an earlier stage are held fixed (but still rendered) while their undecided neighbors are stepped through subsequent stages. All sources in a group are initially modeled as Point Source objects. 
We then determine the appropriate model complexity for each source using a three-stage decision tree, which is controlled by the user-specified tolerances 
$\delta_{\rm SG}$, $\chi^2_{\rm ED}$, $\chi^2_{\rm force, C}$, and 
$\delta_{\rm ED}$. The three stages are:

\begin{enumerate}
    \item Point Source vs. SimpleGalaxy: 
    We first compare the Point Source and SimpleGalaxy models. 
    The Point Source model is preferred if
    \begin{equation}
        \chi^2_{\nu,\rm PS} - \chi^2_{\nu,\rm SG} < \delta_{\rm SG},
    \end{equation}
    i.e., if the improvement from using SimpleGalaxy is smaller than the 
    tolerance $\delta_{\rm SG}$. However, if both models yield 
    $\chi^2_\nu > \chi^2_{\rm ED}$, the source is considered sufficiently 
    extended and is forced to advance to the next stage, regardless of the 
    Point Source preference.

    \item ExpGalaxy vs. DevGalaxy: 
    For sources that advance beyond the first stage, we fit both the 
    ExpGalaxy and DevGalaxy models. A model 
    $X \in \{\rm Exp, Dev\}$ is considered an improvement over 
    SimpleGalaxy if
    \begin{equation}
        \chi^2_{\nu,X} < \chi^2_{\nu,\rm SG}.
    \end{equation}
    The preferred model among Point Source, SimpleGalaxy, 
    ExpGalaxy, and DevGalaxy is then selected as follows:
    \begin{itemize}
        \item If Point Source is preferred over SimpleGalaxy, 
        ExpGalaxy, and DevGalaxy, with the same 
        $\delta_{\rm SG}$ tolerance applied to the comparison with 
        SimpleGalaxy, the source is assigned the Point Source model.

        \item If SimpleGalaxy is preferred over both ExpGalaxy and 
        DevGalaxy, and satisfies 
        $\chi^2_{\nu,\rm SG} < \chi^2_{\rm force,C}$, the source is assigned the 
        SimpleGalaxy model.

        \item If either ExpGalaxy or DevGalaxy clearly outperforms 
        the other, such that
        \begin{equation}
            \left|\chi^2_{\nu,\rm Exp} - \chi^2_{\nu,\rm Dev}\right| 
            > \delta_{\rm ED},
        \end{equation}
        and the preferred model satisfies 
        $\chi^2_\nu < \chi^2_{\rm force,C}$, that model is selected.

        \item Otherwise, if ExpGalaxy and DevGalaxy provide 
        comparable fits or if neither provides a satisfactory fit, the source is 
        promoted to the final stage.
    \end{itemize}

    \item Composite galaxy: 
    In the final stage, we fit the FixedCompositeGalaxy model and compare it with the ExpGalaxy and DevGalaxy models. FixedCompositeGalaxy model is selected if it has the lowest 
    $\chi^2_\nu$ among the three models. Otherwise, the better-fitting model between ExpGalaxy and DevGalaxy is retained, with ties broken in favor 
    of ExpGalaxy.
\end{enumerate}

Based on our simulation tests, we adopt $\delta_{\rm SG}=0.1$, $\chi^2_{\rm ED}=0.15$, $\chi^2_{\rm force,C}=0.15$, and $\delta_{\rm ED}=0.1$, as these values provide the best overall performance in terms of magnitude recovery, detection purity, and completeness. We therefore fix to these values throughout the paper. 

\subsubsection{Photometry measurement}
After model types have been assigned to all sources in the group, we perform a joint optimization across the Y106, J129, H158, F184, and K213 bands\footnote{We have also tested optimization only over Y106, J129, and H158 and found that it doesn't change the paper's conclusion.}, while holding the model type of each source fixed. This step refines the positions and morphological parameters of all sources simultaneously, ensuring that blended sources are modeled self-consistently. 

The resulting catalog of model types, positions, and morphological parameters is then used as input to the multi-band forced-photometry step. In this step, the position and morphological parameters of each source are held fixed, while only its flux is allowed to vary independently in each band. Fixing the source model reduces the number of free parameters in each fit and avoids potential overfitting that could arise if the morphology were re-estimated separately in every band. To reduce degeneracies among neighboring sources fitted within the same group, we impose a positivity prior on the fluxes, $\mathrm{flux}>0$, during the multi-band forced-photometry step. We find that this prior has no noticeable effect on the measured fluxes, but improves the runtime by approximately a factor of two.  The final output is a catalog of fluxes for Y106, J129, H158, F184, and K213 for each detected source, with corresponding uncertainties quantified using the methods described in the following section.

\begin{figure*}
    \centering
    \includegraphics[width=1.0\linewidth]{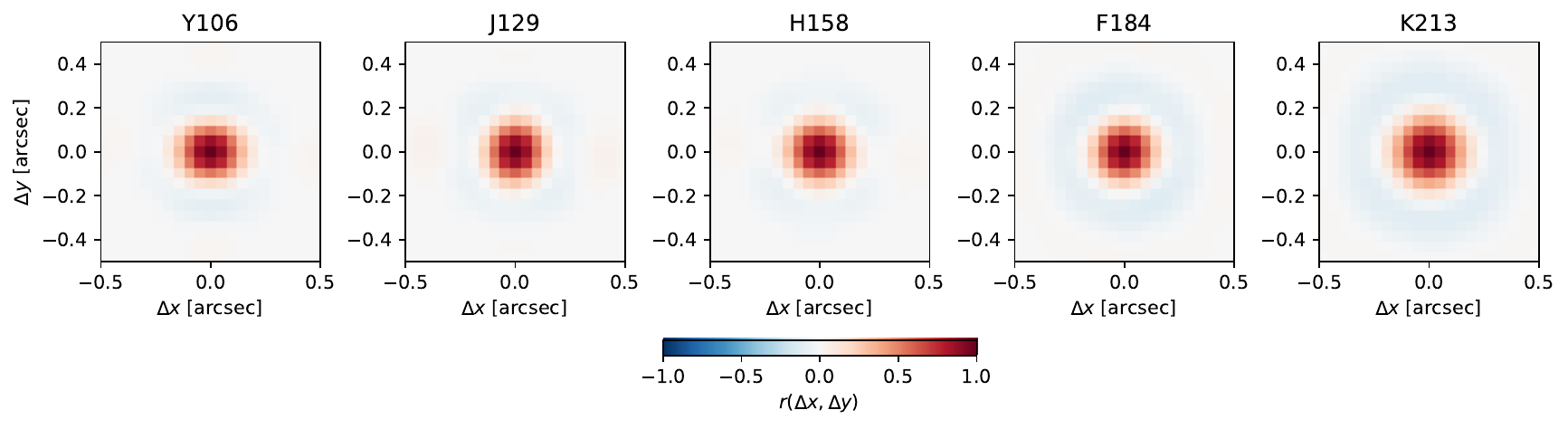}
    \caption{Two-dimensional correlation functions of the normalized noise fields, defined in equation~\ref{eq:corr}, for the Y106, J129, H158, F184, and K213 bands. The correlation functions are computed from a single \pyimcom{} block. Any spatial extent beyond a delta function at zero lag indicates the presence of correlated noise.}
    \label{fig:corr}
\end{figure*}

\subsection{Photometric uncertainties}

In this section, we describe the \texttt{slimfarmer} error estimate, with particular attention to the case of correlated noise.
To simplify the discussions below, let's first rewrite the main likelihood defined in equation \ref{eq:mainlikelihood} as 
\begin{eqnarray}
\label{eq:chi2}
    \chi^2 &=&     \sum_{b \in \rm{bands}}
 \chi^2_{b}, \nonumber \\
    \chi^2_{b}&=& w_b\left(d_{b}-\sum_{i\in g} f_{i,b} T_{i,b} \right)^2, \nonumber\\
    T_{i,b}&=& \hat{m}_{i,b}(\theta)\ast\rm{PSF}_b , 
\end{eqnarray}
where $f_{i,b}$ is the flux of galaxy $i$ in band $b$, and $\hat{m}_{i,b}$ is the normalized model. In the multi-band forced-photometry step, we hold the model type, position, and shape parameters fixed and optimize only the source fluxes. For source $i$ in band $b$, 
minimizing equation~\ref{eq:chi2} gives the inverse-variance-weighted flux estimate
\begin{eqnarray}
    \hat{f}_{i,b} = 
    \frac{\sum_\alpha w_{b,\alpha}\, T_{i,b,\alpha}\, d_{i,b,\alpha}}
    {\sum_\alpha w_{b,\alpha}\, T_{i,b,\alpha}^2},
\end{eqnarray}
where $\alpha$ runs over pixels in the group, and $d_{i,b,\alpha}$ is the image after subtracting the best-fit models of all other sources in the group.
\subsubsection{Photometry Uncertainty with uncorrelated noise}

If the noise is uncorrelated and the per-pixel variance is correctly described by \(1/w_\alpha\), the variance of the flux estimator \(\hat{f}_{i,b}\) is
\begin{equation}
\label{eq:tractorerr}
    \mathrm{Var}(\hat{f}_{i,b}) = \frac{1}{\sum_\alpha w_{b,\alpha} T_{i,b,\alpha}^2},
\end{equation}
consistent with \texttt{Tractor} \citep[Eq.~10 of][]{Farmer}. This expression assumes both that \(\mathrm{Var}(d_{i,b\alpha})=1/w_{b,\alpha}\) and that the model provides an adequate description of the data. In practice, these assumptions may not hold exactly because of observational systematics or imperfections in the noise model. To account for such mismatches, we rescale the flux variance using the residual map. Specifically, we multiply the variance by the reduced chi-square,
\begin{equation}
    \chi^2/\mathrm{dof}, 
\end{equation}
where \(\chi^2\) is computed from the residual map and
\begin{equation}
    \mathrm{dof} = n_{\rm footprint\_pixels} - n_{\rm params},
\end{equation}
$n_{\rm footprint\_pixels}$ is the number of pixels of $d_{b, \alpha}$ and $n_{\rm params}$ is the number of parameters of the fitted model.
This rescaling is exact when the assumed noise model is correct up to an overall multiplicative factor and provides an empirical correction when the model variance underestimates or overestimates the observed residual scatter.

\subsubsection{Photometry Uncertainty with correlated noise}
In practice, however, the noise can be correlated. Such correlations can arise from both the 
\pyimcom\ image-combination procedure and the readout process of Roman's Teledyne H4RG 
detectors \citep{noisecorrelation}. The detector-induced correlations are expected to be largely mitigated by the Improved Roman Reference Correction (B. J. Rauscher et al., in preparation; Roman technical note Roman-STScI-000673, SE-01 by Betti et al.) and \texttt{imdestripe} \citep{imdestripe}, whereas correlations introduced by the image-combination procedure remain. We discuss the contributions from these different 
sources in appendix~\ref{app:noise_correlation}. Because ignoring correlated noise can lead to underestimated flux uncertainties \citep{Masci2009,2014MNRAS.439....2V}, we derive a modified expression for the flux variance below.

Assuming inverse-variance weights, the most general form for the variance of \(\hat{f}_{i,b}\) is
\begin{equation}
    \mathrm{Var}(\hat{f}_{i,b}) =
    \frac{\sum_{\alpha,\beta}
    \sqrt{w_{b,\alpha}}\,T_{i,b,\alpha\,}
    r(\alpha,\beta)\,
    \sqrt{w_{b,\beta}}\,T_{i,b,\beta}}
    {\left(\sum_\alpha w_{b,\alpha} T_{i,b,\alpha}^2\right)^2},
\end{equation}
where \(r(\alpha,\beta)\) is the correlation function of the normalized noise field and $\alpha$, $\beta$ are pixel indices.

Directly estimating \(r(\alpha,\beta)\) is noisy when only a limited number of noise realizations are available. We therefore assume that, within each $1.66'\times1.66'$ block, the correlation function depends only on pixel separation, rather than on the absolute positions of the two pixels \footnote{While the empirical tests in section~\ref{sec:absolutephotometyandcolor} demonstrate the validity of this correction, we note that we have not proved that this assumption holds in general. Developing a thorough understanding of the noise covariance in reconstructed images and the validity of this assumption is left for future work.}. Writing a pixel index as $\alpha=(x,y)$, this assumption gives
\begin{equation}
\label{eq:corr}
    r(\alpha,\beta)
    =
    \left\langle
    \bar{n}(x+\Delta x,y+\Delta y)\,
    \bar{n}(x,y)
    \right\rangle,
\end{equation}
where the normalized noise field is defined as
\begin{equation}
    \bar{n}(x,y) = \sqrt{w_{x,y}}\,n(x,y).
\end{equation}
We estimate \(r\) from the de-biasing noise fields \(n(x,y)\) provided by \pyimcom{} (defined in section \ref{sec:sim}). Figure~\ref{fig:corr} shows an example of the resulting correlation function, computed from a single \pyimcom{} block.

In principle, the model-fitting likelihoods, such as equations \ref{eq:mainlikelihood} and \ref{eq:model_selection}, should also include corrections for correlated noise. In practice, however, our estimate of the noise correlations depends on the fitted galaxy models. Incorporating these correlations directly into the fitting would therefore require an iterative procedure, substantially increasing the computational cost. Because we find that the current implementation yields acceptable photometric performance without including correlated noise during the model-fitting stage, we leave this extension to future work.

\subsection{Forced photometry}
\label{sec:forcedphotometry}

We use simulated Rubin images to test the forced-photometry pipeline. The Rubin images are organized into $2.8~\deg^2$ tracts, each subdivided into $10\times10$ equally sized patches. For each Roman block, we identify all Rubin patches that overlap the block, including an additional $6''$ margin on each side. For each group of galaxies detected in the Roman data, we then fit the corresponding model to the Rubin images using the Rubin PSF models, WCS solutions, and the likelihood defined in Eq.~\ref{eq:mainlikelihood}. In this forced-photometry fit, we vary only the flux and position parameters, while fixing all other model parameters to their best-fit values from the Roman data. Following \texttt{The Farmer} \citep{Farmer}, we also impose a Gaussian position prior with a width of $0.1''$ to account for possible astrometric uncertainties between the Roman and Rubin images.
\subsection{Boundary effect and stellar mask}
Adjacent Roman blocks each extend by $34$ pixels beyond their nominal boundaries. Consequently, the $+34$-pixel boundary region of one block overlaps the $-34$-pixel boundary region of the adjacent block, producing a total overlap of $68$ pixels. The extension on each side corresponds to $1.67''$. To ensure that boundary effects do not affect our analysis, we enlarge this overlap region to $6''$ by loading inner parts of adjacent blocks. We perform detection and photometry on the extended footprint, but include in the final catalog only sources whose positions lie within the central, non-overlapping region of each block. In this way, we ensure that the sources in the central block are not affected by the block's artificial boundaries and avoid double-counting sources near the boundary. 

Finally, we want to mask out diffraction spike features caused by saturated stars. In this paper, we use the true star catalog provided in the \ou{} simulation. We mask out pixels around each star with $m_{\rm{g}}<23$ with a magnitude-dependent radius defined in equation 7 of \cite{y6shearcatalog}.
\begin{figure*}
    \centering
    \includegraphics[width=1.0\linewidth]{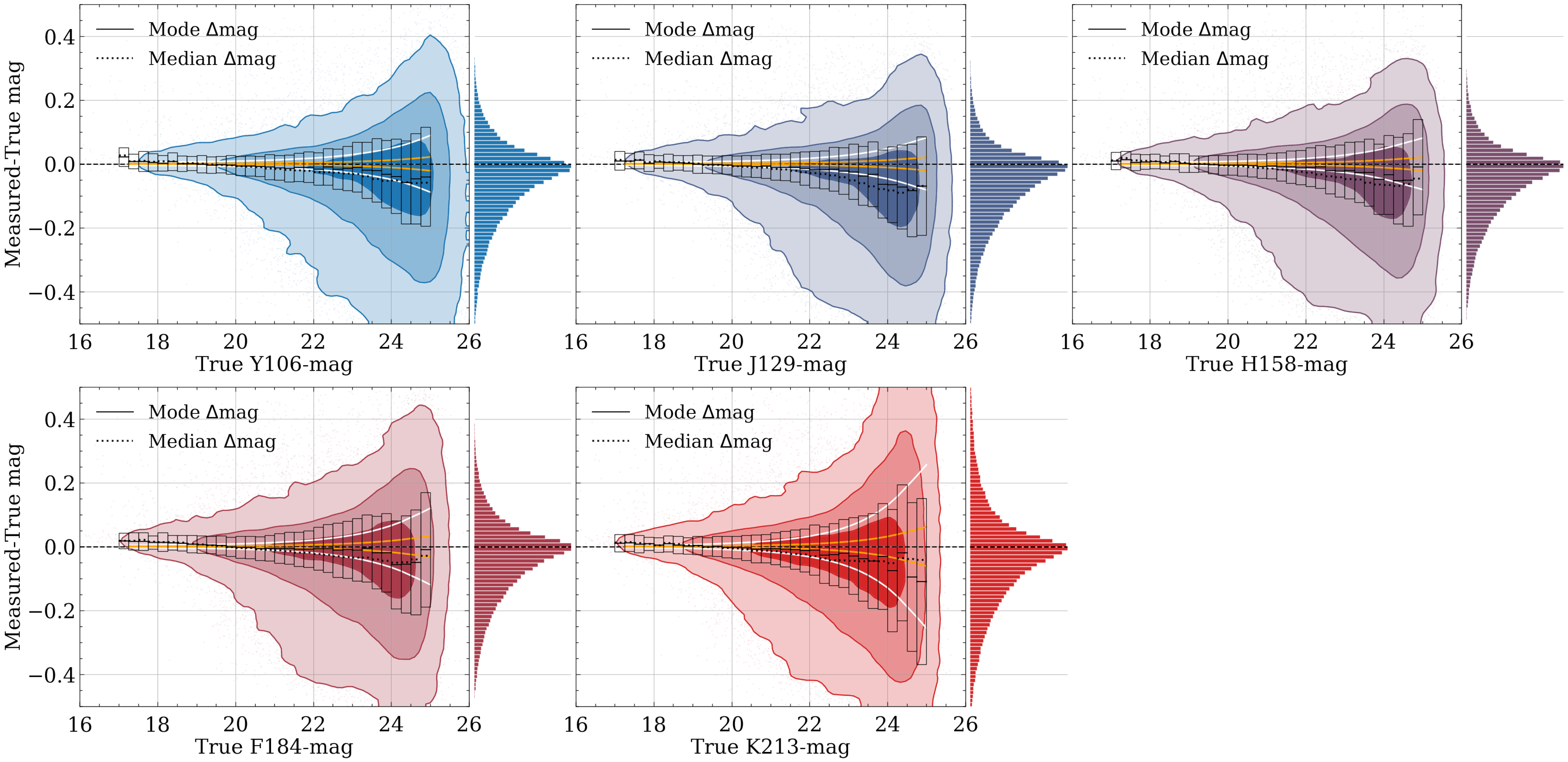}
    \caption{Distribution of the difference between the measured galaxy AB magnitude and the true galaxy AB magnitude as a function of true AB magnitude for the galaxy sample in Y106, J129, H158, F184, and K213 bands. The contours enclose $39.2\%$, $86.5\%$, and $98.9\%$ of the data, corresponding to the $1\sigma$, $2\sigma$, and $3\sigma$ regions of a two-dimensional Gaussian, respectively. We also show the median (dotted line), mode (solid line), and standard deviation (box) of the magnitude difference. The yellow curves show the average magnitude uncertainty in each bin, based on \texttt{The Tractor}'s out-of-the-box estimations (equation \ref{eq:tractorerr}). The white curves show the average magnitude uncertainty in each bin, as estimated by the \slimfarmer{}. }
    \label{fig:mag}
\end{figure*}

\section{Accuracy and Precision of Photometry Measurement} 
\label{sec:absolutephotometyandcolor}
Having performed detection and photometry on \Dcsim{}, we next assess the accuracy and precision of the photometric measurements. Before diving into these validation tests, we first clarify how the photometry will be used in the planned Roman High-Latitude Imaging Survey Project Infrastructure Team (HLIS-PIT)'s cosmological analyses. The relevant requirements depend on the downstream science application, and identifying these use cases allows us to optimize the photometric measurements accordingly. In particular, Roman photometry will be used to select galaxies and galaxy cluster samples for clustering and weak-lensing analyses, assign galaxies to tomographic redshift bins, and estimate the redshift distribution of each bin. In this paper, we pay particular attention to applications related to galaxy clustering, galaxy cluster sample selections, and photometric redshift characterization. These applications require reliable measurements of galaxy colors and color uncertainties, with performance that is well characterized and minimally correlated with observational properties such as depth, crowding, and observing conditions\footnote{While the dependence of photometry performance on observing conditions is important for cosmology, we do not have the product to test this and therefore have to defer it for future work.}.

To quantify the photometry performance, we first define a representative source-galaxy sample by selecting objects with a signal-to-noise ratio greater than $18$ in the Y106, J129, H158 mean combined detection image \citep{tim2021}. This selection yields a galaxy number density of $37.8~\mathrm{arcmin}^{-2}$. We then match the detected galaxies to the truth catalog using the spatial-plus-magnitude (S+M) matching algorithm described in \citet{dc1}. Specifically, for each detected galaxy, we first identify all truth objects within $0.147''$, corresponding to three pixels in the simulated image. Among these candidates, we select the truth object whose five-band magnitude vector, defined using the Y106, J129, H158, F184, and K213 bands, has the smallest Euclidean distance from the measured magnitudes. We require this five-band magnitude distance to be less than $1.0$. With this procedure, $96.3\%$ of the sample is matched to a truth galaxy, consistent with the matching performance reported in \citet{dc1}. Among the remaining unmatched detections, about $73\%$ have a truth galaxy within $0.147''$ but fail the magnitude-matching criterion, with a five-band magnitude distance greater than $1.0$. These cases likely correspond to photometric failures or severe blending. The remaining $27\%$ have no truth galaxy within $0.147''$, which are primarily associated with spurious detections due to imperfect stellar masks. We show example cutouts of these failure cases in figure~\ref{fig:bad}.
We also find that only $0.004\%$ of detected galaxies are matched to the same truth object as another detection. These duplicate matches typically occur when bright galaxies are shredded into multiple detected objects, again similar to the behavior found in \cite{dc1}.

\subsection{Photometry recovery}
During our initial analysis of \Dcsim{}, we found that the background-subtraction step in our pipeline can oversubtract galaxy flux, potentially biasing the photometric validation. The oversubtraction is around $0.02$ mag at $m_{\rm H}=18$ and $0.13$ mag at $m_{\rm H}=24.5$. Since background subtraction is not the focus of this paper, we isolate the performance of the detection and photometry pipeline by using a simplified observed image. Specifically, rather than using the fully end-to-end processed coadds, we use the true coadded image with the de-biasing field added. This image retains the relevant noise properties, including background noise and astronomical shot noise from galaxies, while avoiding biases introduced by imperfect background subtraction. 

We show the difference in measured photometry (in AB magnitude) and true photometry (in AB magnitude) as a function of true photometry in Roman bands in figure~\ref{fig:mag} and table~\ref{tab:photometry_accuracy}. The density contours enclose $39.3\%$, $86.5\%$, and $98.9\%$ of the sample, corresponding to the $1\sigma$, $2\sigma$, and $3\sigma$ regions of a two-dimensional Gaussian, respectively. In bins of
width $0.25$ mag, we overplot the mean and mode biases, and scatter of measured and true magnitude differences as dotted lines, solid lines, and black boxes, respectively. The scatter is estimated from the 16th and 84th percentiles
to reduce sensitivity to outliers. The white curves show the $1\sigma$ magnitude uncertainty estimated in \slimfarmer{} while the yellow curves show the out-of-the-box magnitude uncertainty from \texttt{The Tractor}.  We find that the magnitude uncertainty estimated by \texttt{The Tractor} significantly underestimates the empirical magnitude uncertainty due to the lack of quantification of correlated noise. The \slimfarmer{}, on the other hand, is more consistent with the empirical estimates of magnitude uncertainty.  Similar to the finding in \cite{y6balrog}, the uncertainty estimates are slightly smaller than the empirical estimations. This is expected because the empirically estimated scatter includes additional effects, such as depth variations. Such a discrepancy is larger for bright objects when the additional effects dominate the magnitude differences, and smaller for faint objects when the noise dominates the source of scatter. In addition, consistent with \citet{dc1}, we find that the magnitude residuals are small for bright objects but become increasingly negative toward the faint end. In appendix~\ref{app:magnitude_residual}, we show that this trend is unlikely to be driven by selection effects near the detection threshold. Instead, the bias is more likely driven by flux contamination from nearby detected and undetected sources, including unmodeled light from galaxies below the detection threshold and residual deblending errors in multi-galaxy groups.

\begin{figure*}
    \centering
\includegraphics[width=0.75\linewidth]{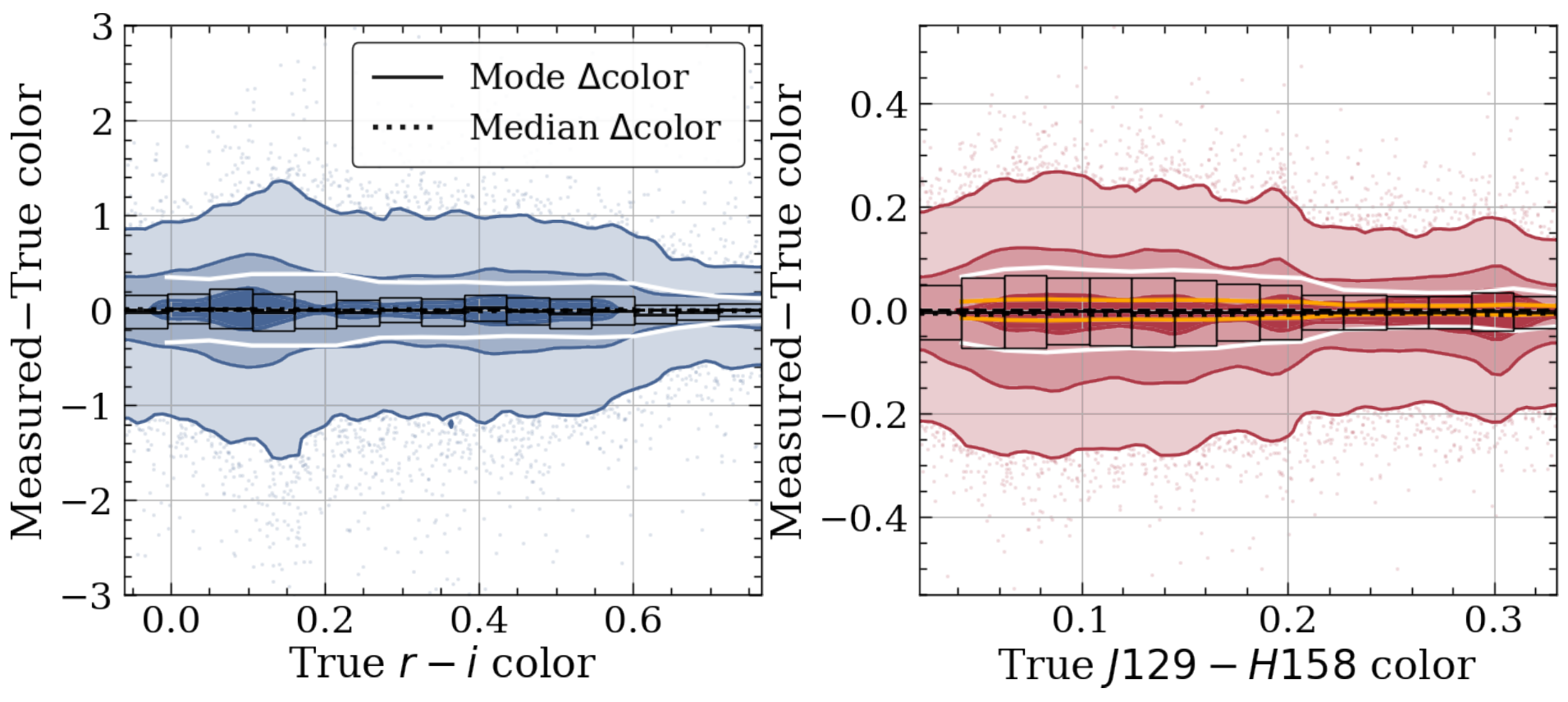}
    \caption{
Same as figure~\ref{fig:mag}, but for the differences between the measured and true colors in Rubin and Roman bands. We show the $\rm{r}-\rm{i}$ and $\mathrm{J129}-\mathrm{H158}$ colors as representative Rubin-based and Roman-based colors, respectively. The summary statistics are computed in 15 bins from the $10$ to the $90$ percentiles of the color distribution of source galaxy samples (S/N $>18$ in the mean combined detection image). The \slimfarmer{}-based color uncertainty, computed by adding the corresponding magnitude uncertainties in quadrature, agrees well with the empirical estimate, while the out-of-the-box \texttt{The Tractor}-based color uncertainty is significantly smaller than the empirical estimate. The full set of color comparisons is shown in figure~\ref{fig:color}.
    }
    \label{fig:color_short}
\end{figure*}
\begin{figure*}
    \centering
\includegraphics[width=0.75\linewidth]{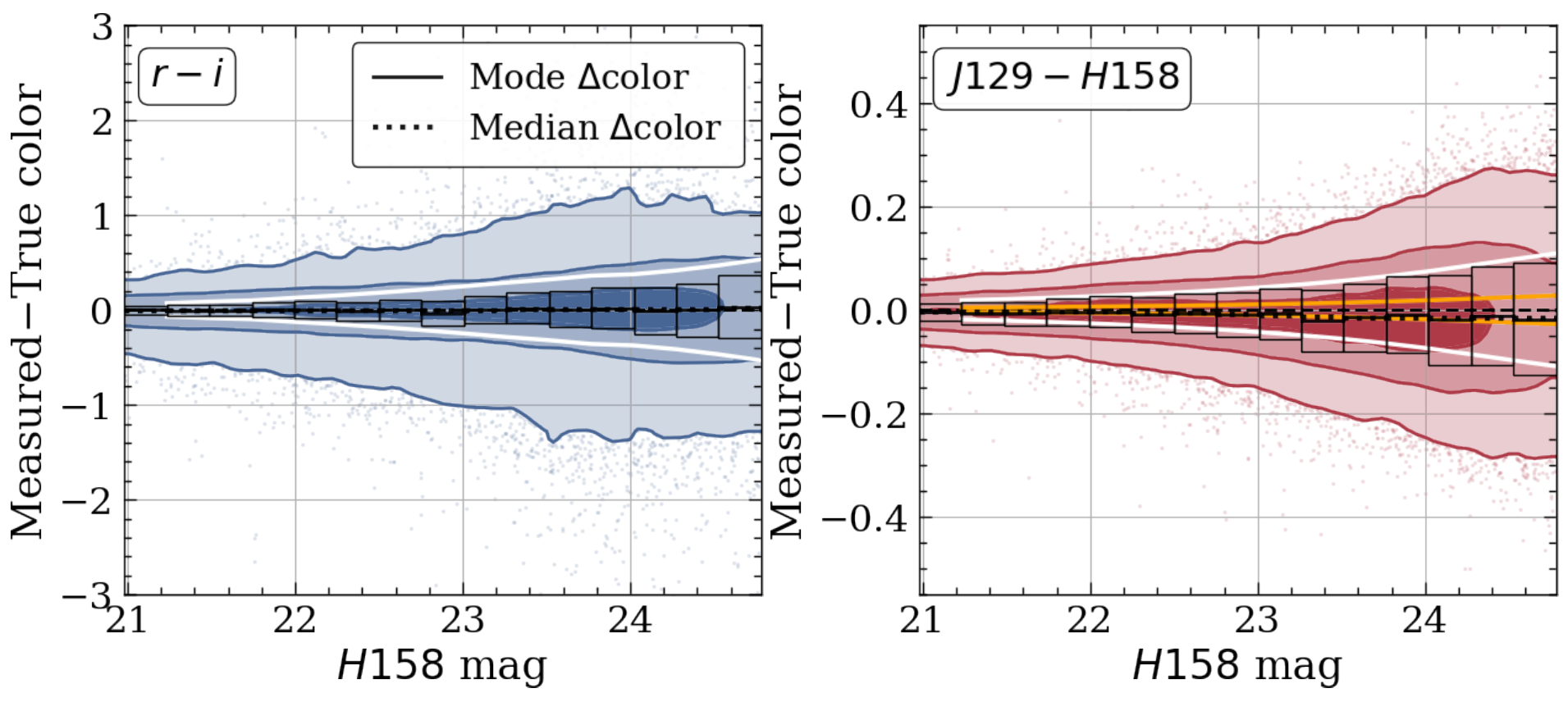}
    \caption{
Same as figure~\ref{fig:color_short}, but for the differences between the measured and true colors in Rubin and Roman bands as a function of H158 magnitude. The summary statistics are computed in 15 bins from the $10$ to the $90$ percentiles of the H158 magnitude distribution of source galaxy samples (S/N$>18$ in the mean combined detection image). The yellow curves show the average color uncertainty in each bin, based on \texttt{The Tractor}'s out-of-the-box estimations (equation \ref{eq:tractorerr}). The white curves show the average color uncertainty in each bin, as estimated by the \slimfarmer{}.
    }
    \label{fig:colormagnitude_short}
\end{figure*}

\subsection{Color recovery}

We now validate the performance of the color measurements. Because both Roman and Rubin colors will be used in the planned cosmological analyses, we include both surveys in this test. 

\subsubsection{Color recovery vs galaxy properties}

Figures~\ref{fig:color_short} and ~\ref{fig:color} show the difference between the measured and true galaxy colors for the Rubin u,g,r,i,z,y bands and the Roman Y106, J129, H158, F184, and K213 bands. The density contours and summary statistics are computed in the same way as in figure~\ref{fig:mag}, using the same galaxy sample selection.

We find that the color residuals are substantially more consistent with zero than the individual magnitude residuals. In addition, the uncertainties estimated by \slimfarmer{} agree much better with the empirical scatter for colors than for magnitudes, consistent with the findings of \citet{y6balrog}. Quantitatively, the mode residual is within $20$ millimag for all Roman-based colors and within $70$ millimag for Rubin-based colors. This improved behavior likely reflects the fact that several effects that bias observed magnitudes or increase the empirical scatter are correlated across bands and therefore partially cancel in color measurements. For example, near the detection threshold, selection on the mean-combined Roman detection image can induce correlated flux residuals across different bands, while depth variations can similarly coherently affect multiple bands. As in the magnitude validation, we find that the out-of-the-box \texttt{The Tractor} uncertainties significantly underestimate the empirically measured color scatter. This highlights the importance of modeling correlated noise in order to obtain reliable color uncertainties. We also find that the color scatter is substantially larger in the Rubin bands than in the Roman bands. This is expected because we do not impose a signal-to-noise cut in the Rubin bands; as a result, the Rubin forced photometry has low signal-to-noise (SNR) for a large fraction of the Roman-detected sample (with a median SNR=2.2, 5.7, 6.0, 4.7, 3.8, 2.9 in u g r i z y bands respectively). We note that, as described in section~\ref{sec:forcedphotometry}, the simulated Rubin images correspond to the five-year depth of Rubin's LSST, which is shallower than the Rubin data expected for the final Roman--Rubin analyses.  

Figures~\ref{fig:colormagnitude_short} and \ref{fig:colormagnitude} show the color residuals as a function of H158 magnitude. The residuals remain consistent with zero across the magnitude range for the Roman-based colors and most of the Rubin-based colors. However, we find that the color residuals show deviations at the faint end ($\rm{mag}_{\rm H158}>25$) in several Rubin-based colors, such as z-y and y-Y106. These deviations likely reflect the particularly low signal-to-noise ratio of the Rubin forced photometry for these faint, Roman-selected galaxies, which can lead to photometric biases. 
In section~\ref{sec:colorvscrowdedness}, we find no significant dependence of these residuals on galaxy crowdedness, while the tests in section~\ref{sec:ensembleredshift} provide no evidence that they degrade photometric-redshift characterization. We therefore defer a more detailed investigation of the origin and mitigation strategy of these faint-end deviations to future work.

\subsubsection{Color recovery vs crowdedness}
\label{sec:colorvscrowdedness}
\begin{figure*}
    \centering
    \includegraphics[width=1.0\linewidth]{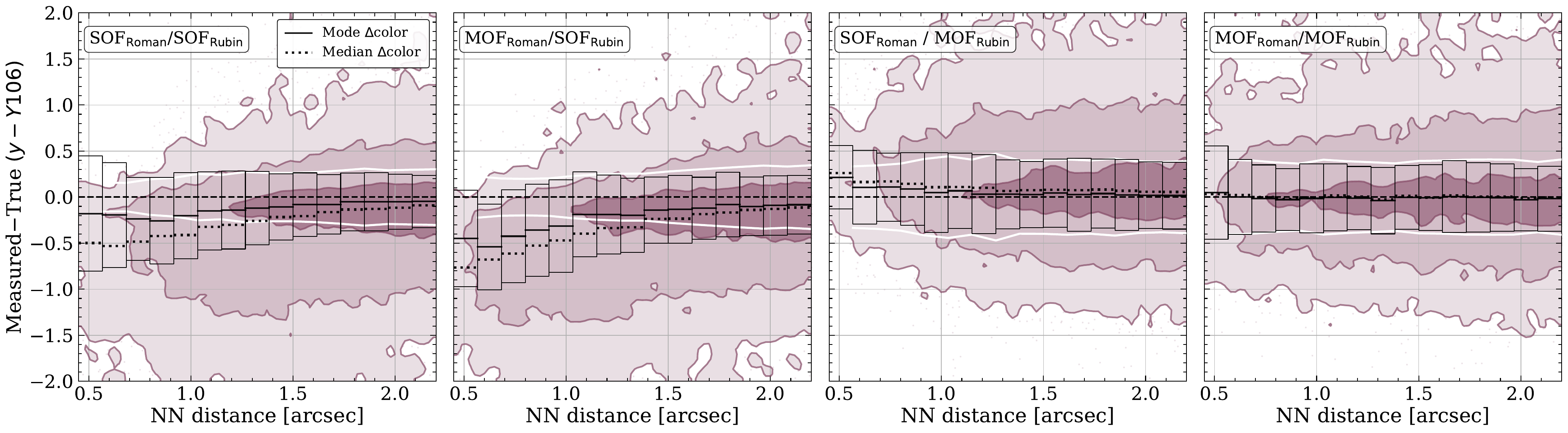}
    \caption{
The plot configuration is the same as in Figure~\ref{fig:color_short}, but here we show the differences between the measured and true colors in the Rubin y and Roman Y106 bands as a function of distance to the nearest neighbor. The summary statistics are computed in 15 equally spaced bins spanning $0.45''$ to $2.2''$. The last panel shows the default \slimfarmer{} configuration, while the remaining panels compare different combinations of multi-object and single-object fitting in the Roman and Rubin bands. The corresponding results for other bands, using the default \slimfarmer{} configuration, are shown in figure~\ref{fig:color_nndistance}; the same configuration comparison as shown here, but for the i-H158 color, is shown in figure~\ref{fig:color_nndistance_iH}.}
    \label{fig:color_nndistance_different}
\end{figure*}
Because galaxy colors are the primary properties used to select galaxies whose clustering properties are of great interest, we want to ensure that the color measurement does not introduce an artificial clustering signal. Specifically, we do not want the color measurement performance to depend on the clustering properties of galaxies. One way to quantify this is to compare the measured and true galaxy color as a function of distance to nearest neighbors. In the last panel of figure~\ref{fig:color_nndistance_different}, we show such a comparison in the Rubin y band and the Roman Y106 band as a representative case using the default \slimfarmer{} configuration. We chose the y-Y106 color as a representative case because colors that combine one Rubin band and one Roman band are expected to be especially sensitive to differences in resolution between the two surveys. We have also validated the i-H158 color (figure~\ref{fig:color_nndistance_iH}) and find results consistent with figure~\ref{fig:color_nndistance_different}. For other colors measured from adjacent bands, we show them in figure~\ref{fig:color_nndistance}. We find that \slimfarmer{}, when run in the default configuration, delivers consistent performance of galaxy color measurements and error quantification regardless of the crowdedness of galaxy environments (all the way down to $0.45''$, which is about $2\times$ the Roman PSF FWHM) in Rubin- and Roman-based color.

\subsubsection{Color recovery vs photometry algorithms}
Having validated the fiducial \slimfarmer{} photometry, we next test the sensitivity of the results to one of the main pipeline choices: whether blended sources are fit jointly or independently. In this work, we use multi-object fitting (MOF) to refer to our \slimfarmer{} configuration in which all sources in a group defined by overlapping dilated segmentation maps are modeled simultaneously. We use single-object fitting (SOF) to refer to the alternative configuration in which each source is fit independently, while neighboring sources are not jointly optimized. Similar distinctions between joint and independent source fitting have been explored in the Dark Energy Survey (DES) photometric catalogs \citep[e.g.,][]{y1gold,y3gold}, although the implementations differ. This comparison is particularly important for Roman--Rubin colors, because the two surveys have different PSFs and therefore different blending properties. Fitting blended sources independently can assign neighbor flux differently in the Roman and Rubin bands, producing biased colors. We therefore focus this test on color residuals as a function of nearest-neighbor distance, which directly probes whether the photometry introduces environment-dependent color biases relevant to photometric redshift estimation.
There are four possible combinations: MOF or SOF in the Roman bands, crossed with MOF or SOF in the Rubin bands. In the MOF-Roman/SOF-Rubin configuration, galaxies are modeled jointly in the Roman bands, but each galaxy is fit independently during Rubin forced photometry. As shown in the second panel of figure~\ref{fig:color_nndistance_different}, the color bias becomes increasingly blue in crowded environments, reaching approximately $0.5$ mag at a nearest-neighbor separation of $0.5''$. This occurs because neighboring objects are not jointly modeled in the Rubin bands; as a result, blending can make the Rubin fluxes too bright, shifting the Roman--Rubin colors blue. We also test the opposite configuration, SOF-Roman/MOF-Rubin, in which Roman models and fluxes are fit independently, while Rubin forced photometry is performed jointly for groups of galaxies with overlapping dilated segmentation maps. In this case, the colors become biased red in crowded regions, with the bias reaching approximately $0.2$ mag at a nearest-neighbor separation of $0.5''$. This bias arises because the Roman fluxes can include contributions from blended neighbors that are not jointly modeled. This effect is smaller than in the MOF-Roman/SOF-Rubin case. This reflects that Rubin has a larger PSF and therefore suffers more from blending. Finally, in the SOF-Roman/SOF-Rubin configuration, we again find a blue bias in crowded environments. We find that multi-object fitting is necessary for both Roman and Rubin to achieve crowdedness-independent color-measurement performance. 
 
In summary, these validation tests show that \slimfarmer{} provides robust
photometry for the HLIS-PIT cosmological analyses. Although the observed
photometry exhibits some biases at the faint end, partly due to selection
effects near the detection threshold, galaxy colors are recovered with
substantially smaller biases. The \slimfarmer{} uncertainty estimates also
track the empirical color scatter much more accurately than the default
\texttt{The Tractor} uncertainties, demonstrating the importance of accounting
for correlated noise. We further find that color performance remains stable as
a function of nearest-neighbor distance when multi-object fitting is applied
consistently in both the Roman and Rubin bands. In contrast, using
single-object fitting in either survey induces environment-dependent color
biases. These results motivate our default choice of multi-object fitting for both Roman photometry and Rubin forced photometry, as it provides an effective way to mitigate blending-induced color systematics in this photometry-focused validation. For weak-lensing applications, however, the impact of this choice on shear measurements must also be considered; we leave this investigation to future work. 
\section{Impacts on photometric redshift estimates}
\label{sec:totalredshift}
With the photometric measurements in hand, we now investigate whether the
resulting color measurements are sufficiently well behaved for downstream
photometric-redshift (photo-$z$) calibration. When discussing weak-lensing applications in this section, we focus specifically on source redshift characterization, rather than on shear calibration or the coupling between shear measurement and redshift calibration. Broadly, two types of photo-$z$
performance are relevant for HLIS-PIT cosmological analyses, and more generally
for weak lensing, galaxy clustering, and cluster cosmology analyses \citep[e.g.,][for a
review]{photozreview}. First, weak-lensing analyses require accurate characterization of the redshift
distributions of galaxies assigned to a small number of tomographic bins. In
this case, the key questions are how well separated the tomographic bins are
and how biased the inferred redshift distributions may be, especially when the
redshift calibration sample is not fully representative of the weak-lensing
source sample. Second, galaxy-clustering and cluster-finding analyses rely more directly on
redshift estimates for individual galaxies. In this case, the main concerns are
the scatter and outlier rate of the estimated photo-$z$ values at fixed true
redshift, as well as any dependence of these errors on galaxy environment. In this section, we therefore perform two representative photo-$z$ tests,
targeting these two use cases, and quantifying how the performance of
\slimfarmer{} propagates into photo-$z$ calibration.

\subsection{Ensemble photometric redshift}
\label{sec:ensembleredshift}
\begin{figure}
    \centering
    \includegraphics[width=0.8\linewidth]{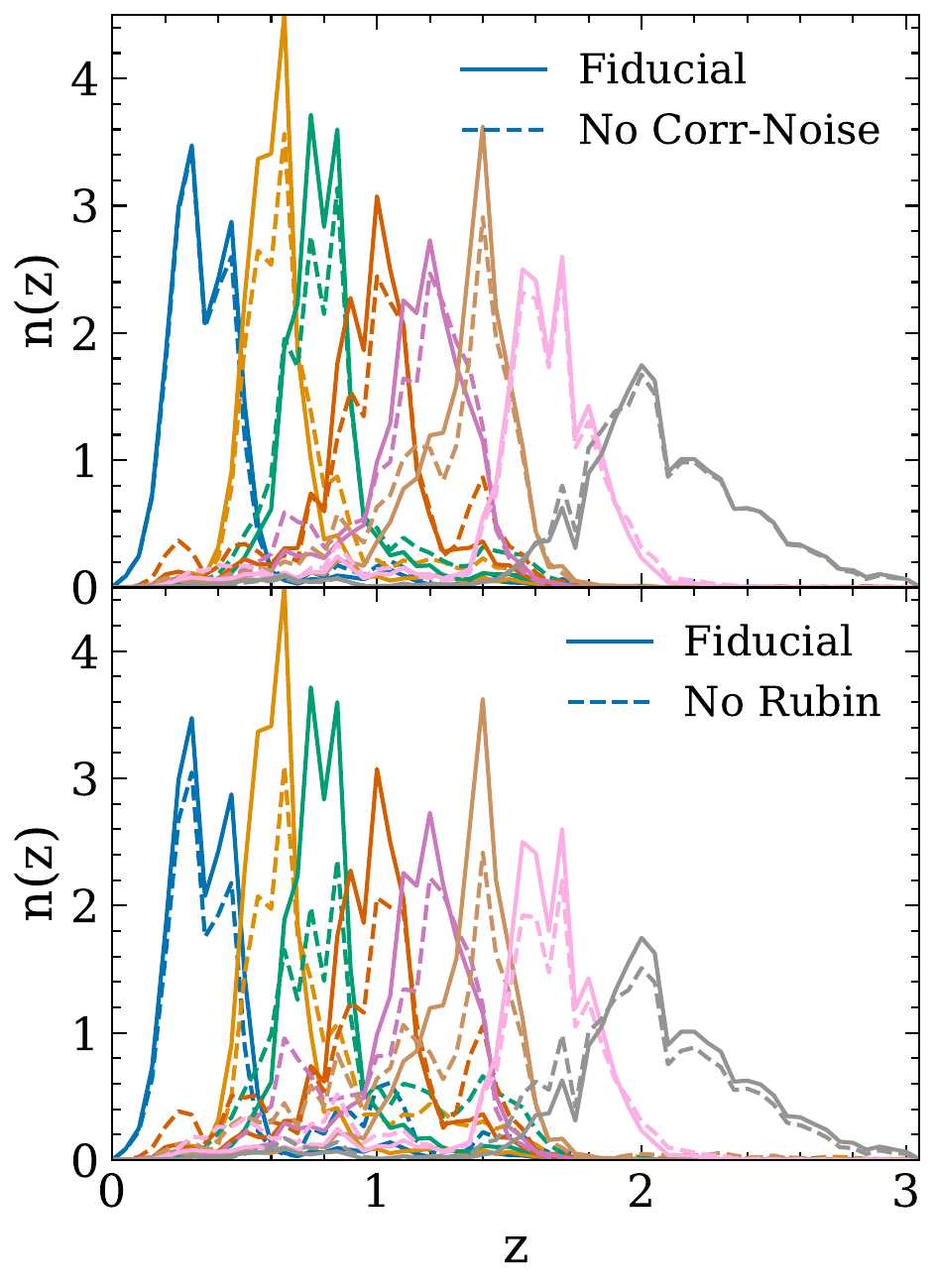}
    \caption{
Redshift distributions estimated by \texttt{Roman SOMPZ} for simulated source galaxies selected with $\mathrm{S/N}>18$ in the mean-combined detection map. In both panels, solid lines show the Fiducial configuration described in Section~\ref{sec:ensembleredshift}, which uses Roman Y106, J129, and H158 photometry together with Rubin u, g, r, i, z, and y photometry, measured with the default \slimfarmer{} configuration. The dashed lines show two alternative configurations: the top panel uses the out-of-the-box \texttt{The Tractor} photometric uncertainties without our correlated-noise correction, while the bottom panel excludes the Rubin photometry. In both cases, the inferred redshift distributions exhibit substantially larger overlap between tomographic bins than in the Fiducial configuration.
    }
    \label{fig:sompzmain}
\end{figure}

For the photo-$z$ method that characterizes the redshift distribution of an ensemble of galaxy samples, we adopt the self-organizing-map (SOM)-based photometric-redshift calibration method (SOMPZ), following the approach used in the DES Year 3 and Year 6 cosmological analyses
\citep{SOMPZ-DES_Y3,y6somdata}. A self-organizing map (SOM) is an unsupervised machine-learning algorithm that projects the high-dimensional space of galaxy photometry onto a two-dimensional grid, such that galaxies with similar multi-band colors are assigned to the same or nearby cells, often referred to as phenotypes. SOMPZ uses the redshift calibration sample to estimate the redshift distribution associated with each phenotype. These distributions are then combined according to the phenotype occupations of the source sample to construct the ensemble $n(z)$.

This basic idea is straightforward when the source sample and the redshift
calibration sample are described by the same photometric information. In
Roman HLIS, however, the relevant samples can have different photometric depth
and wavelength coverage \citep{timeallocation}. The wide or medium-tier sample defines the galaxy
population whose redshift distribution is needed for cosmological analyses, but
it has more limited photometric information. The deep-tier sample provides more
informative multi-band photometry that helps break color--redshift
degeneracies, while only a much smaller subset of galaxies has spectroscopic
redshifts. SOMPZ therefore provides a natural framework for connecting these
samples: a wide SOM describes the source population, a deep SOM provides a higher-fidelity characterization of phenotypes that have a narrow redshift distribution, and a transfer
function maps galaxies between the two SOMs. This allows redshift information
calibrated in the deep photometric space to be propagated to the wide-field
source sample.

In this analysis, we run the \texttt{Roman SOMPZ} pipeline, described in
appendix~\ref{app:sompz}, using the measured \slimfarmer{} photometry. The
pipeline follows the DES SOMPZ methodology but is implemented in a modular
framework using \texttt{RAIL} v1.2.4 \citep{rail} and \texttt{ceci} v2.3.1\footnote{https://github.com/LSSTDESC/ceci}, allowing the different stages
of the analysis and their nuisance parameters to be specified through a common
configuration file. Since the goal of this section is to assess the impact of \slimfarmer{} photometry on photo-$z$ calibration, we adopt a simplified SOMPZ setup in which the same photometric measurements are used for both the deep and wide SOM assignments. 

In principle, the \slimfarmer{} measurements can be evaluated through two complementary photo-$z$ diagnostics. The first concerns how photometric quality affects the sensitivity of the inferred $n(z)$ to incompleteness in the redshift-calibration sample. If the photometry separates galaxies into phenotypes with very narrow redshift distributions, then even a biased calibration sample can recover an unbiased $n(z)$; in the limiting case where each phenotype has a delta-function redshift distribution, only one calibration galaxy per phenotype would be needed. Degraded photometry, by contrast, broadens the redshift distribution within each phenotype and makes the final $n(z)$ more sensitive to missing or non-representative calibration galaxies. The second diagnostic is how photometric quality affects tomographic binning: that is, whether the measured colors allow galaxies to be assigned to redshift bins that are well separated in true redshift. Testing the first diagnostic would require constructing a realistic mock redshift calibration sample, including its selection function and incompleteness, and could be prone to simulation artifacts, such as the width of redshift--broadband color relations. We therefore focus on the second diagnostic and test how the \slimfarmer{} photometry affects our ability to define tomographic redshift bins.

To isolate the impact of photometric measurements on redshift binning, we assume that $30\%$ of the source galaxies have spectroscopic redshifts and that these galaxies constitute a representative subsample of the full source-galaxy sample. By construction, this removes biases associated with spectroscopic selection and allows us to focus specifically on how different photometric configurations affect the SOMPZ-based photometric redshift binning. We consider the following five scenarios:
\begin{enumerate}
    \item \textbf{Fiducial:} Our fiducial case includes the Roman Y106, J129, and H158 bands, together with the Rubin u, g, r, i, z, and y bands. We use the default \slimfarmer{} configuration, which performs multi-object fitting for both Roman and Rubin photometry. This setup is designed to mimic the Roman medium-tier analysis.

    \item \textbf{No Corr-Noise:} We pass the out-of-the-box \texttt{The Tractor} photometric uncertainties directly to SOMPZ, without applying our correlated-noise correction.

    \item \textbf{Roman SOF/ Rubin MOF:} We use single-object fitting (SOF) for Roman photometry and multi-object fitting (MOF) for Rubin photometry.

    \item \textbf{Roman SOF/ Rubin SOF:} We use single-object fitting (SOF) for both Roman and Rubin photometry.

    \item \textbf{No Rubin:} We ignore the Rubin bands and use only the Roman photometry. The No Rubin configuration is included only as a limiting case to quantify the impact of Rubin optical photometry on photometric-redshift characterization; it is not intended to represent a realistic baseline for Roman HLIS cosmology analyses.
\end{enumerate}
In all five cases, galaxies are selected using a signal-to-noise threshold of $\mathrm{S/N}>18$ measured from the mean combined map, ensuring an apples-to-apples comparison across the different photometric configurations. We define the redshift bin edges using the true redshift of each galaxy and partition the galaxies into eight equal-number-density bins.  

\begin{figure*}
    \centering
    \includegraphics[width=1.0\linewidth]{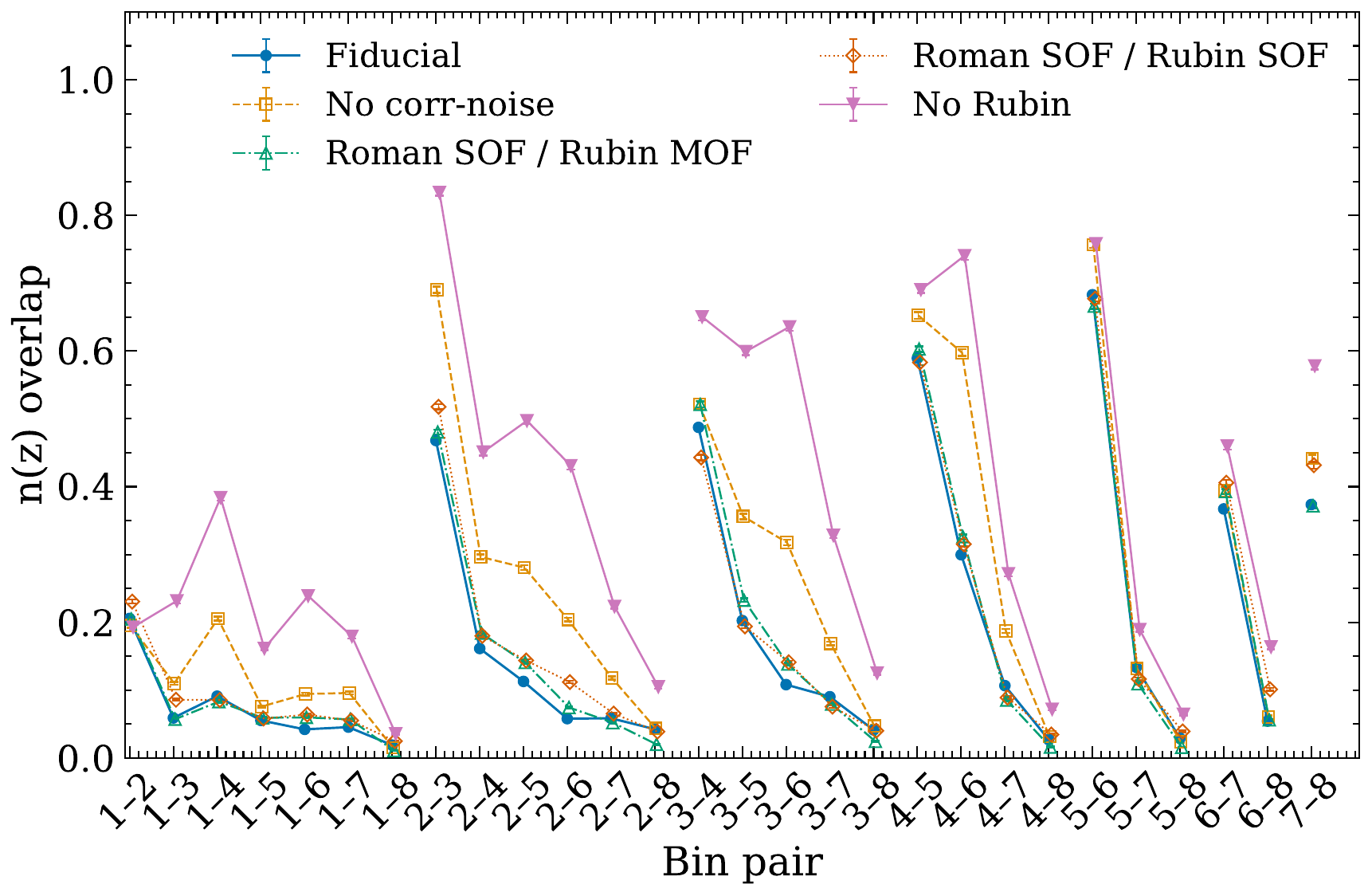}
    \caption{Redshift-bin overlaps for the Fiducial configuration and the alternative photometric configurations described in section~\ref{sec:ensembleredshift}, including the No Corr-Noise, Roman SOF/Rubin MOF, Roman SOF/Rubin SOF, and No Rubin cases. The overlap is quantified using the normalized Gram matrix of the inferred redshift distributions, with larger values indicating greater overlap between bins. Error bars are estimated from realizations of $n(z)$ assuming sample variance for a $20\ \deg^2$ field, comparable to the Roman deep-tier area.
    }
    \label{fig:sompzoverlap}
\end{figure*}
Figure~\ref{fig:sompzmain} shows the redshift distributions inferred by
\texttt{Roman SOMPZ} for galaxies assigned to each tomographic redshift bin in
the Fiducial, No Corr-Noise, and No Rubin cases. Ignoring correlated noise
broadens the inferred redshift distributions and increases the overlap between
tomographic bins. Removing the Rubin bands leads to an even stronger
degradation, substantially reducing the separation between bins. To quantify this effect, we compute the overlap between tomographic redshift
bins following \citet{y6som}. Specifically, we calculate the normalized Gram
matrix of the redshift distributions, defined for bins $p$ and $q$ as
\begin{equation}
O_{p,q}
=
\frac{\sum_z n_p(z)\,n_q(z)}
{\sqrt{\sum_z n_p^2(z)}\sqrt{\sum_z n_q^2(z)}} .
\end{equation}
Here, $O_{p,q}=1$ corresponds to identical redshift distributions,
while $O_{p,q}=0$ indicates no overlap. The resulting overlap
measurements are shown in figure~\ref{fig:sompzoverlap}. The error bars are estimated from the realizations of
$n(z)$, assuming sample variance corresponding to a $20\,\deg^2$ field, roughly the size of the Roman deep tier.

As expected, the $n(z)$ overlap decreases with bin-pair separation, and across most bin pairs, the Fiducial configuration yields the smallest overlap. Removing the Rubin bands significantly increases the overlap, highlighting the importance of optical photometry for Roman tomographic binning. This result is particularly notable because the \ou{} simulations have substantially narrower infrared colors at fixed redshift than the COSMOS2020 data \citep{cosmos20}, making the simulated infrared colors more predictive of redshift than they are likely to be in real data and therefore making the No Rubin case artificially favorable. Even under this optimistic assumption, we still find that the Rubin bands are essential for robust tomographic bin assignment. We also find that correcting for correlated noise is important: using the uncorrected \texttt{The Tractor} uncertainties leads to systematically larger bin overlap. This is likely because the color uncertainties enter the metric used to determine distances between galaxies in the high-dimensional color space. Incorrect uncertainty estimates, therefore, distort this metric and degrade the performance of the SOM. By comparison, the choice between SOF and MOF photometry for Roman and/or Rubin has only a minor impact on the bin-overlap metric.
\begin{figure*}
    \centering
    \includegraphics[width=0.8\linewidth]{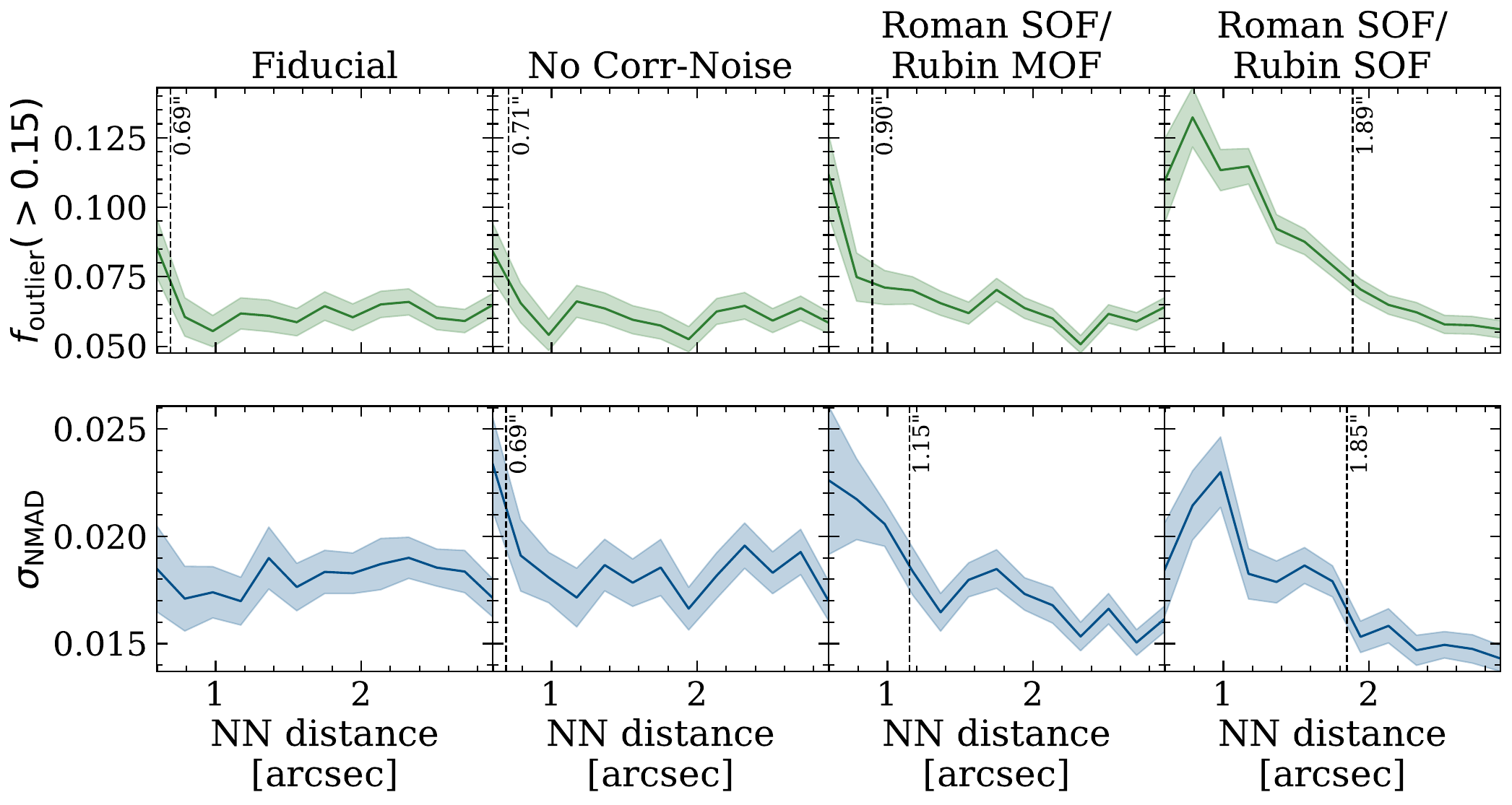}
    \caption{Individual photo-$z$ performance as a function of nearest-neighbor separation, quantified by the outlier fraction (top row) and the median absolute deviation (bottom row). Statistics are measured in 15 equally spaced bins between $0.5''$ and $3''$. Columns correspond to the photometric configurations described in Section~\ref{sec:ensembleredshift}: Fiducial, No Corr-Noise, Roman SOF/Rubin MOF, and Roman SOF/Rubin SOF. Vertical lines mark the nearest-neighbor separation at which each curve first deviates by more than $2\sigma$ from the median values, where both the median and $\sigma$ are estimated using sources with nearest-neighbor separations larger than $2''$. Error bars show the $1\sigma$ uncertainties estimated from $1000$ bootstrap resamplings. The No Rubin case is omitted for visual clarity, because its outlier fraction ($\sim 0.2$) and median absolute deviation ($\sim 0.04$) are substantially larger than those of the other configurations. 
    }
    \label{fig:individual_photoz}
\end{figure*}
\subsection{Individual photometric redshift}
For the photo-$z$ method that estimates the redshift of individual galaxies, we adopt the \texttt{Flexzboost} algorithm \citep{2017arXiv170408095I,2020A&C....3000362D} implemented in \texttt{Rail} \citep{rail}, which uses gradient boosted decision trees to predict photo-$z$ Posterior Distribution Functions (PDF) by minimizing the Conditional Density Estimate loss. This choice should not be interpreted as indicating that the HLIS-PIT has selected \texttt{FlexZBoost} as the baseline point-estimate photo-$z$ algorithm. Rather, we adopt it because it can be applied flexibly to simulations, unlike many template-fitting methods that require simulation-specific template construction, and because it has demonstrated good performance in LSST Data Preview 1 analyses \citep{2026MNRAS.tmp.1049Z}.

We run \texttt{Flexzboost} for all five scenarios presented in section~\ref{sec:ensembleredshift}. Again, we assume that the spectroscopic galaxies form a representative subset of the source galaxy sample to isolate the impact of photometric measurements. The galaxies are selected using a signal-to-noise threshold $>18$ measured from the mean-combined map. To quantify the individual photo-$z$ performance, we measure the following quantity, 
\begin{eqnarray}
    \Delta z &=& \frac{z_{p}-z_{s}}{1+z_s} \\
    f_{\rm outlier} &=& N(|\Delta z|>0.15) / N\\
    \sigma_{\rm NMAD} &=& 1.48\times \rm{median}(|\Delta z-median \left(\Delta z\right)|), 
\end{eqnarray}
where $z_{s}$ is the spectroscopic redshift, $z_{p}$ is the \texttt{FlexZBoost} photo-$z$, and N is the total number of objects. One primary goal is to ensure both  $f_{\rm outlier}$ and $\sigma_{\rm NMAD}$ are as small as possible and that the values do not depend on galaxy environments.

Figure~\ref{fig:individual_photoz} compares the outlier fraction, $f_{\rm outlier}$, and the normalized median absolute deviation, $\sigma_{\rm NMAD}$, as functions of nearest-neighbor separation for the photometric configurations defined in section~\ref{sec:ensembleredshift}, with error bars estimated from $1000$ bootstrap resamplings. The No Rubin case is not shown for visual clarity, because its outlier fraction ($\sim 0.2$) and median absolute deviation ($\sim 0.04$) are substantially larger than those of the other configurations. We find that the Fiducial configuration shows little dependence of the photo-$z$ statistics on local galaxy environment, except in the most crowded regions, whereas the Roman SOF/Rubin MOF and Roman SOF/Rubin SOF configurations exhibit stronger environmental trends. Ignoring the correlated-noise correction has minimal impact on the \texttt{FlexZBoost} photo-$z$ performance. To quantify the scale at which crowding begins to affect the photo-$z$ statistics, we define the transition scale as the first nearest-neighbor separation, moving inward from $3''$, at which a curve differs from its reference median by more than $2\sigma$. For each curve, both the reference median and $\sigma$ are estimated using sources with nearest-neighbor separations larger than $2''$.  With this definition, the transition scale for Roman SOF/Rubin MOF is approximately twice that of the Fiducial configuration, while that for Roman SOF/Rubin SOF is approximately three times larger. Interestingly, the Roman SOF/Rubin MOF and Roman SOF/Rubin SOF configurations yield slightly lower $\sigma_{\rm NMAD}$ and $f_{\rm outlier}$ in less crowded environments, possibly because these measurements have slightly lower photometric noise than the MOF/MOF measurements, as shown in figure~\ref{fig:color_nndistance_different}. However, we caution that the overall normalization of these curves is sensitive to the number of spectroscopic calibration samples. In the results above, we assume that $30\%$ of galaxies have spectroscopic redshifts; reducing this fraction to $10\%$ can raise the large-separation outlier fraction to $\sim 0.1$ and the median absolute deviation to $\sim 0.035$ for the Fiducial configuration. This suggests that the large-separation normalization of $f_{\rm outlier}$ and $\sigma_{\rm NMAD}$ is dominated by the availability of spectroscopic calibration data, while the relative environmental trends are more directly sensitive to photometric performance.

Finally, although we find that the performance of \texttt{FlexZBoost} is only minimally affected in the No Corr-Noise scenario, this result may not generalize to other photometric-redshift algorithms. In particular, representative spectroscopic-redshift samples may be limited during the early stages of Roman HLIS analyses, motivating the use of template-based photo-$z$ methods such as \texttt{LePhare} \citep{Lephare}. Because these methods explicitly use photometric uncertainties to construct the likelihood, inaccurate uncertainty estimates could bias the inferred redshift probability distributions or lead to overconfident constraints. A robust evaluation with the present simulations would require constructing an appropriate template library capable of representing the galaxy spectral energy distributions used to generate the simulated images. We therefore defer an assessment of the impact of photometric-uncertainty calibration on template-based photo-$z$ methods to future work.

\section{Conclusions}
\label{sec:conclude}
In this work, we develop and validate a photometry measurement pipeline, \slimfarmer{}, for the Roman High-Latitude Imaging Survey. The pipeline performs source detection on Roman coadded images, groups neighboring galaxies using dilated segmentation maps, and measures multi-band photometry through multi-object model fitting with \texttt{The Tractor}. We apply this framework to the \texttt{DC25} Roman image simulations, which are processed through a substantial subset of the High-Latitude Imaging Survey Project Infrastructure Team (HLIS-PIT) image-processing pipeline. We also perform Rubin forced photometry on matched \ou{} Rubin image simulations, allowing us to test the joint Roman--Rubin photometry that will be needed for photometric-redshift characterization in Roman HLIS cosmology analyses. For weak-lensing applications, the discussion here focuses on source-galaxy redshift characterization. Shear calibration and the coupling between shear calibration and redshift characterization are beyond the scope of this work. We list our main findings below:

\begin{enumerate}
\item  Multi-object model fitting provides precise and accurate photometric measurements for both Roman photometry and Rubin forced photometry (section~\ref{sec:absolutephotometyandcolor}). The color measurements are robust: the mode of the color residuals is within $20$ millimag for Roman-based colors and within $70$ millimag for Rubin-based colors. The observed photometry agrees well with the input truth catalog, although it shows faint-end biases, partly due to blending with detected and undetected neighboring galaxies (appendix~\ref{app:magnitude_residual}).

\item Accounting for correlated noise in Roman coadded images is important for reliable photometric uncertainties (figures~\ref{fig:mag} and ~\ref{fig:color_short}). In the simulations considered here, correlated noise arises from both detector readout and image combination, whose relative importance is discussed in appendix~\ref{app:noise_correlation}. We find that the out-of-the-box \texttt{The Tractor} uncertainties substantially underestimate the empirical scatter in both observed magnitudes and colors, because they do not account for these noise correlations. In \slimfarmer{}, we derive flux uncertainties that incorporate correlated noise in the coadded images using noise realizations propagated through the same image-processing steps as the science images. The resulting \slimfarmer{} uncertainty estimates agree much better with the empirical scatter for both galaxy colors and observed magnitudes.

\item We adopt the model-fitting approach to model the astronomical shot noise, which is a significant fraction of the total error budget of Roman HLIS's weak lensing sample. In appendix~\ref{app:shot_noise} and figure~\ref{fig:shotnoise}, we explore the impact of different modeling choices on the accuracy and precision of the photometry measurement. We find that using the science image itself to model astronomical shot noise will significantly bias the photometry measurement. 

\item For the photometric fluxes and colors tested here, consistent multi-object fitting in both Roman and Rubin is needed to avoid environment-dependent color biases (figure~\ref{fig:color_nndistance_different}). We compare configurations in which Roman and Rubin photometry are measured with either multi-object fitting (MOF) or single-object fitting (SOF). When Rubin forced photometry is performed with SOF, blended Rubin fluxes are biased high in crowded regions, shifting Roman--Rubin colors blue. Conversely, when Roman photometry is measured with SOF, but Rubin photometry is measured with MOF, Roman fluxes can include blended neighbor light, producing redder colors in crowded environments. These results show that consistent MOF photometry in both Roman and Rubin is needed to obtain color measurements that are stable with the galaxy environment.

\item Rubin optical photometry is essential for Roman tomographic redshift binning (figures~\ref{fig:sompzmain} and ~\ref{fig:sompzoverlap}). 
Using the \texttt{Roman SOMPZ} pipeline (detailed in appendix \ref{app:sompz}), we find that the joint Roman and Rubin photometry produces the smallest overlap between tomographic redshift bins. Removing Rubin photometry substantially increases the bin overlap, even though the \ou{} simulations likely make the infrared-only case artificially favorable because of their narrow infrared color--redshift relations. This highlights the importance of Rubin optical photometry for Roman ensemble redshift calibration.

\item Environment-dependent photometric biases propagate into individual photo-$z$ performance, as shown in figure~\ref{fig:individual_photoz}. Using \texttt{FlexZBoost}, we find that the configuration using multi-object fitting (MOF) in both the Roman and Rubin bands shows little dependence of the outlier fraction and median absolute deviation on local galaxy environment, except in the most crowded regions. In contrast, configurations that use single-object fitting (SOF), including Roman SOF/Rubin MOF and Roman SOF/Rubin SOF, show stronger environmental trends.
\end{enumerate}

Taken together, these results establish \texttt{slimfarmer} as a prototype photometric framework for Roman HLIS-PIT cosmology applications that require accurate galaxy fluxes, colors, and photometric uncertainties. In the simulations considered here, robust Roman--Rubin color measurements require correlated-noise corrections, a model-based treatment of astronomical shot noise, and consistent multi-object fitting across the Roman and Rubin images. The Rubin optical bands are essential for achieving well-separated tomographic redshift bins, while multi-object fitting is necessary to control environment-dependent color biases and the photo-$z$ systematics they induce. These findings identify the key photometric ingredients that must be included, validated, or propagated in Roman galaxy-clustering and galaxy-cluster analyses, as well as in the source-galaxy redshift characterization needed for weak-lensing cosmology. However, determining the optimal treatment of neighboring sources for weak-lensing cosmology will require validation within the appropriate shear-measurement and shear-calibration framework, since choices about how neighboring flux is modeled or subtracted can bias the resulting shear measurements \citep{2023OJAp....6E..17S}.

Future work will extend this validation in three directions. First, we will test the framework on more realistic simulations that include additional observational and processing systematics, such as imperfect background subtraction, PSF-modeling errors, and time-dependent bandpass effects due to water and ice condensation. Second, we will validate the same methodology on Roman early-observation data using synthetic source injections, following the spirit of similar analyses in DES \citep{y6balrog, y3balrog}, to measure photometric biases and uncertainty calibration directly in real data. Third, we will propagate the photometric systematics quantified in this work into Roman cosmological analyses, including their effects on galaxy and cluster sample selection, redshift-distribution calibration, tomographic-bin overlap, and their downstream impact on cosmic shear, galaxy clustering, galaxy--galaxy lensing, and cluster lensing constraints. For weak-lensing applications, such propagation will need to be combined with an appropriate shear-calibration framework, such as \texttt{METADETECTION} \citep{2020ApJ...902..138S}. 

\textit{Data availability.} The \slimfarmer{} and \texttt{Roman SOMPZ} code, along with the configurations used in this work, will be made publicly available upon acceptance of this publication.

\section*{Acknowledgments}
CHT is supported by Eric and Wendy Schmidt Fellowship. This work was partially supported by NASA grant 22-ROMAN22-0011 Maximizing Cosmological Science with the Roman High-Latitude Imaging Survey Project Infrastructure Team (PIT). This work was completed in part with resources provided by the University of Chicago’s Research Computing Center. CHT thanks the organizers of the Roman-Rubin joint processing workshop, where some of the ideas in the paper were discussed. CHT thanks Dhayaa Anbajagane, Matt Becker, Arun Kannawadi, Charlie Mpetha, Dan Master, Peter Melchior, Jeff Newman, Dan Taranu, Alex Drlica-Wagner, and Masa Yamamoto for helpful discussions on various aspects of this work. This paper makes use of LSST Science Pipelines software developed by the Vera C. Rubin Observatory. We thank the Rubin Observatory for making their code available as free software at https://pipelines.lsst.io.

\newpage
\bibliographystyle{aasjournal}

\bibliography{oja_template}

\begin{appendix}

\section{Weight Map}
\label{app:weight}
In this appendix, we detail how the inverse variance weight is derived from the available PyIMCOM product. 
Here we follow the notation of \cite{Kailiimcom}. The \pyimcom\ image is written as
\begin{equation}
    H_{\alpha} = \sum_{i=0}^{n-1} T_{\alpha i} I_i .
\end{equation}
For simplicity, we adopt the Einstein summation convention below.

The variance of $H_\alpha$ is
\begin{equation}
    V_{\alpha,\alpha} = T^i_\alpha T^j_\alpha \langle I_i I_j \rangle .
\end{equation}
Assuming that the input pixels are uncorrelated, we have
\begin{equation}
    \langle I_i I_j \rangle = \sigma_i^2 \delta_{ij},
\end{equation}
where $\delta_{ij}$ is the Kronecker delta. Substituting this into the expression above gives
\begin{equation}
\label{eq:goal}
    V_{\alpha,\alpha} = \left(T^i_\alpha\right)^2 \sigma_i^2 .
\end{equation}

Equation~\ref{eq:goal} is the quantity we ultimately want. We now relate it to the available noise realizations based on the background-only noise field:
\begin{equation}
    N_\alpha = T^i_\alpha n_i,
\end{equation}
where $n_i$ is a Gaussian random field with zero mean and standard deviation $\sigma_i$.

We can then compute
\begin{eqnarray}
    \sum_\alpha N_\alpha^2 &=& \sum_\alpha \left(T_\alpha^i n_i\right)^2 .
\end{eqnarray}
Now assume that $\sigma_i$ is constant, $\sigma_i = \sigma$. Then
\begin{eqnarray}
\label{eq:goal1}
    \sum_\alpha N_\alpha^2
    &=& \sum_\alpha \sum_i \left(T_\alpha^i\right)^2 \sigma^2 \\
    &=& \sum_\alpha S_\alpha \sigma^2 ,
\end{eqnarray}
where we define
\begin{equation}
    S_\alpha \equiv \sum_i \left(T_\alpha^i\right)^2, 
\end{equation}
which can be obtained from the hdu=6 in the astropy convention (starting with zero) of \pyimcom\ products.

Combining Equations~\ref{eq:goal} and \ref{eq:goal1}, we obtain
\begin{equation}
    V_{\alpha,\alpha}
    = \sum_i \left(T^i_\alpha\right)^2
    \frac{\sum_\alpha N_\alpha^2}{\sum_\alpha S_\alpha} .
\end{equation}
The quantity $\sum_\alpha N_\alpha^2$ is estimated from the simulated noise realizations, which do not include the Poisson shot noise from the source itself. It is well known that the detection weight should not depend on the noisy observed flux directly \citep{Bernstein18}. However, the total noise in the data does include a source Poisson contribution. To account for this, we add a Poisson term computed from the model flux.

From \pyimcom, we use the \textsc{EFFCOVER} map to estimate the effective number of exposures contributing to each pixel. Multiplying this by the exposure time per input exposure gives the total effective exposure time. The corresponding Poisson variance is then taken to be the model flux divided by this total exposure time.

Our final weight map is therefore constructed from the inverse of the total variance,
\begin{equation}
    w_\alpha = \left( V_{\alpha\alpha} + \frac{f_{\rm model,\alpha}}{\mathrm{eff\_gain}_\alpha} \right)^{-1},
\end{equation}
where
\begin{equation}
    \mathrm{eff\_gain}_\alpha = \frac{t_{\rm exp,\alpha}^{\rm eff}}{g},
\end{equation}
with $t_{\rm exp,\alpha}^{\rm eff}$ the effective total exposure time and $g$ the detector gain.

\section{Contribution to noise correlations}
\label{app:noise_correlation}
\begin{figure*}
    \centering
    \includegraphics[width=0.7\linewidth]{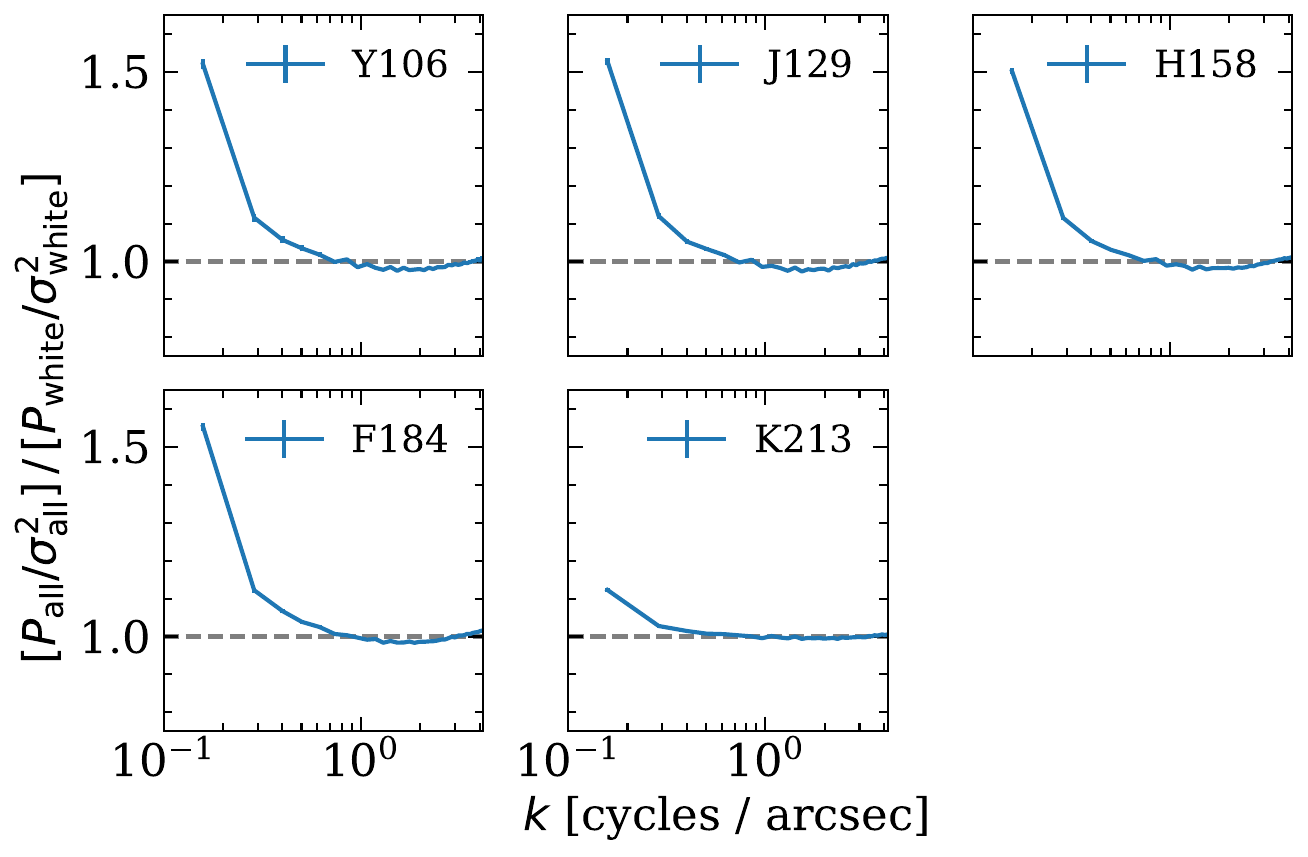}
    \caption{Ratio of the variance-normalized 1D noise power spectrum of the full coadded noise to that of the input white noise. The dot is calculated as the median of $1600$ blocks, and error bars are estimated from $1000$ bootstrap resamples.}
    \label{fig:noisecontribution}
\end{figure*}
In this section, we use \Dcsim{} to understand the sources of noise correlations. As described in Section~\ref{sec:sim}, \Dcsim{} provides two types of noise fields: a white-noise field and a correlated-noise field, each of which is propagated through the \pyimcom{} coaddition process. We first normalize each noise field by the square root of the weight map generated in Appendix~\ref{app:weight} to remove features induced by depth variations. We then compute the variance and the 1D power spectrum of each map (excluding the overlap padding pixels), normalize the power spectra by their respective variances, and take the ratio of the two. Following \citet{noisebias}, we apply a Tukey window before the Fourier transform to suppress unphysical effects from finite sampling. The result is shown in Figure~\ref{fig:noisecontribution}. We find that within the effects included in \Dcsim{}, correlated read noise contributes significantly on large scales, while small scales are dominated by the \pyimcom{} process. The large-scale correlated-noise contribution is reduced in the K213 band because this band is dominated by thermal noise from the telescope\footnote{https://github.com/RomanSpaceTelescope/roman-technical-information}, which dilutes the relative contribution of correlated read noise.

\section{Impact of astronomical shot noise}
\label{app:shot_noise}
\begin{figure*}
    \centering
    \includegraphics[width=0.7\linewidth]{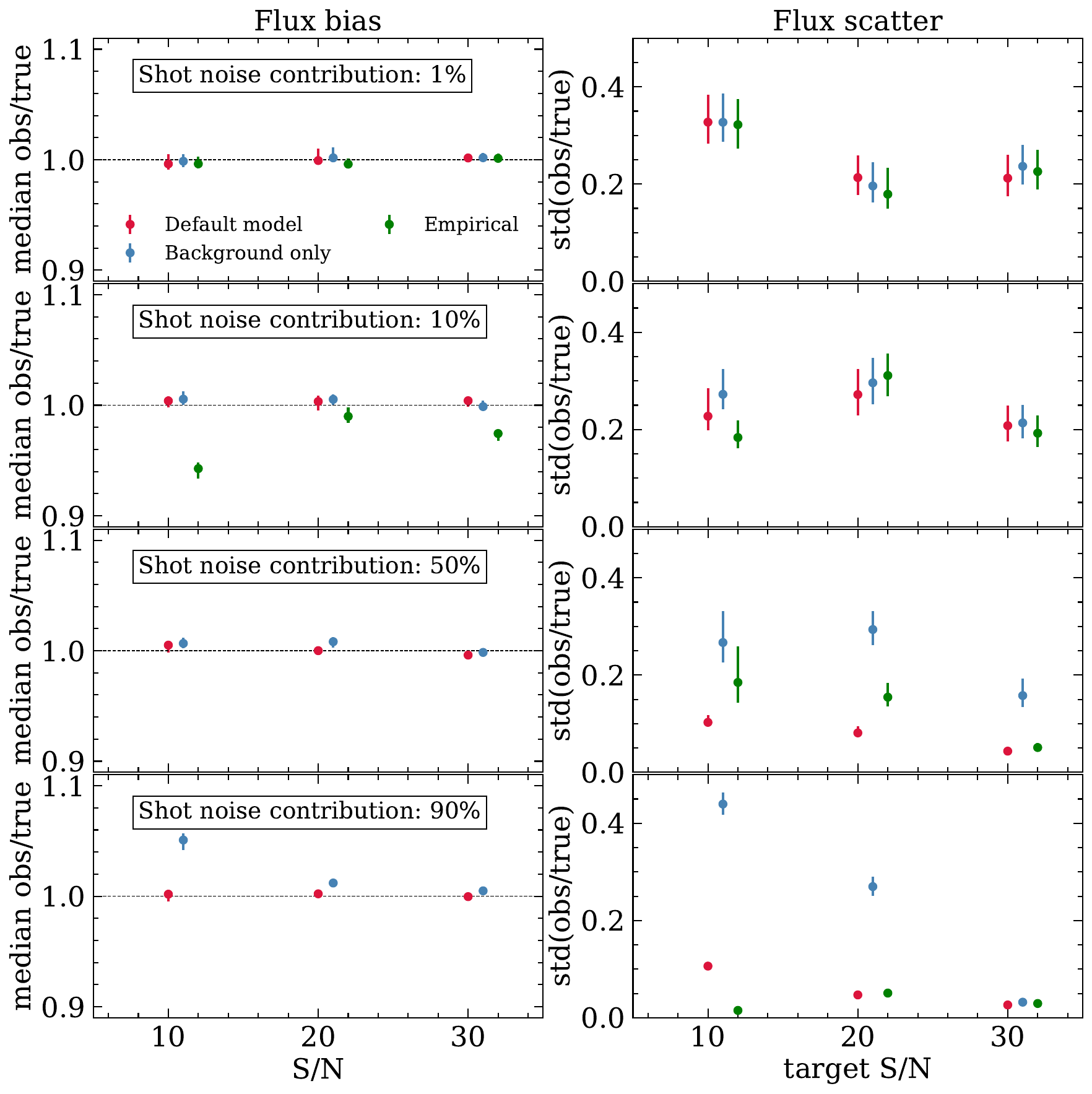}
    \caption{Impact of astronomical shot-noise modeling on flux estimation under different conditions. The first column shows the median ratio of measured flux to true flux, while the second column shows the corresponding standard deviation. Each panel presents these quantities as a function of signal-to-noise ratio, and different rows correspond to different levels of astronomical shot noise. We compare three models for astronomical shot noise: the default \slimfarmer{} model (red), a background-only model that ignores astronomical shot noise (blue), and an empirical model that estimates astronomical shot noise directly from the observed image (green). Points outside the plotting range are not shown. Error bars are estimated using 1000 bootstrap resamplings.} 
    \label{fig:shotnoise}
\end{figure*}
In this appendix, we quantify how the treatment of shot noise from astronomical objects affects model-fitting-based flux estimates. In \slimfarmer{}, we initialize the fit with a background-only weight map and, at each fitting stage, update this map by adding the expected astronomical shot noise predicted by the current model image, as described in appendix~\ref{app:weight}. This model-based treatment is not the only possible choice. For example, \cite{y6shearcatalog} neglects astronomical shot noise during model fitting, while \cite{2023FrASS..1089443T} estimates this contribution directly from the background-subtracted image. It is therefore interesting to learn how these choices affect flux estimates under different conditions.
We compare three treatments of astronomical shot noise:
\begin{enumerate}
    \item \textbf{Default model}: the shot-noise contribution is estimated from the current model image.
    \item \textbf{Background only}: the shot-noise contribution from astronomical objects is ignored.
    \item \textbf{Empirical}: the shot-noise contribution is estimated from the observed image.
\end{enumerate}

We evaluate these choices using controlled image simulations. We consider galaxy flux signal-to-noise ratios of $10$, $20$, and $30$, and vary the fractional contribution of astronomical shot noise to the total noise budget among $1\%$, $10\%$, $50\%$, and $90\%$. For each configuration, we simulate isolated exponential galaxies with an effective size of $0.3''$, approximately the median size of galaxies in \Dcsim{} with $\mathrm{S/N}>18$, and assume a PSF size of $0.24''$. We then run the detection and photometry pipeline on each simulated image and repeat the process $500$ times. The resulting flux biases and scatter are shown in Figure~\ref{fig:shotnoise}.

When the contribution from astronomical shot noise is small ($1\%$), all three treatments produce unbiased flux estimates. This regime is representative of ground-based observations, where the noise budget is often dominated by sky background. However, once the astronomical shot-noise contribution increases to $10\%$, the empirical estimator begins to show measurable biases. At the $50\%$ level, ignoring astronomical shot noise still yields approximately unbiased flux estimates, but with scatter that is several times larger than that of the default model-based treatment. Finally, when astronomical shot noise dominates the noise budget ($90\%$), the background-only treatment also begins to show significant deviations. These results demonstrate that the model-based treatment used in \slimfarmer{} provides more robust flux estimates, especially when source shot noise contributes appreciably to the total noise.

\section{Roman SOMPZ}
\label{app:sompz}
\begin{figure*}
    \centering
    \includegraphics[width=1.0\linewidth]{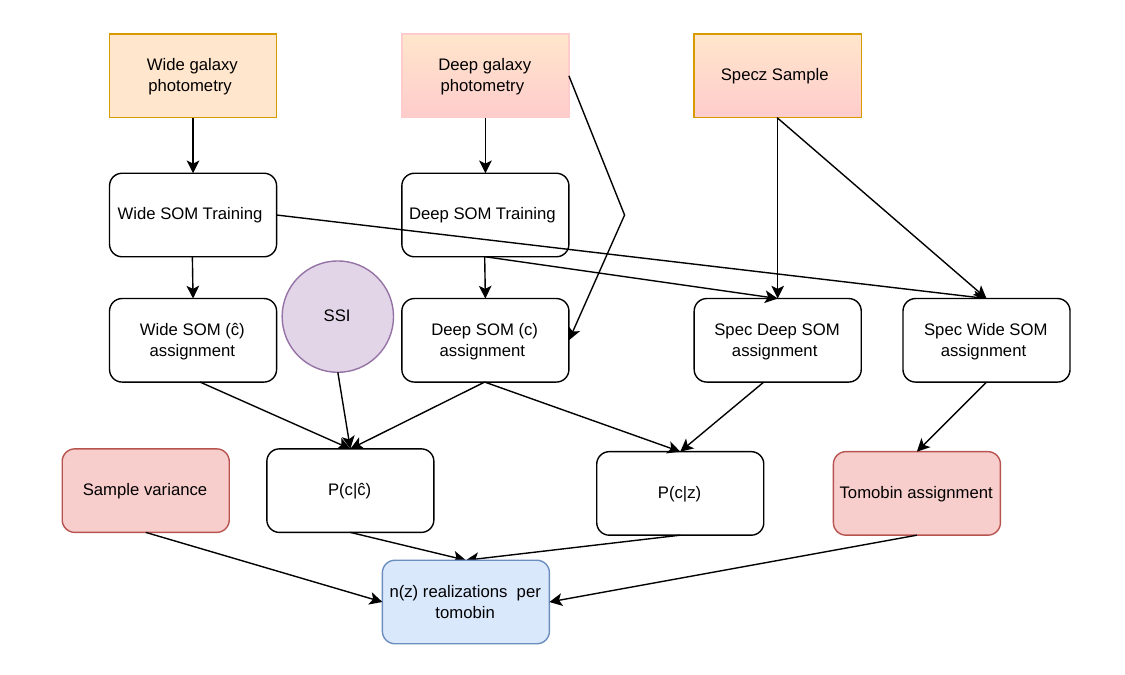}
    \caption{Schematic diagram of the \texttt{Roman SOMPZ} pipeline. Photometric measurements used in the pipeline include wide-galaxy photometry, deep-galaxy photometry, the redshift of the redshift-calibration sample (Specz sample), and synthetic source-injection simulations. The \texttt{Roman SOMPZ} pipeline combines this information with the sample variance calculation to assign galaxies to tomographic bins and to generate realizations of $n(z)$ for each bin, reflecting the best estimates and associated uncertainties.} 
    \label{fig:sompz}
\end{figure*}
This appendix describes the Roman self-organizing-map photometric redshift
(\texttt{Roman SOMPZ}) pipeline, which is publicly available at
\url{https://github.com/Roman-HLIS-Cosmology-PIT/Roman-SOMPZ.git}. The pipeline
follows the methodology developed for DES Year 6 \citep{y6som,y6somdata},
summarized in Fig.~\ref{fig:sompz}, but is implemented in a more modular
framework. In particular, the Roman pipeline uses \texttt{RAIL} v1.2.4\citep{rail} and
\texttt{ceci} v2.3.1 (\url{https://github.com/LSSTDESC/ceci}) to connect the different
stages of the analysis. This structure makes it easier to run the pipeline
end-to-end, replace individual components, and specify nuisance parameters for
each stage through a common YAML configuration file. Below, we briefly describe
the \texttt{Roman SOMPZ} framework.

SOMPZ uses self-organizing maps (SOMs) to partition the high-dimensional space of
galaxy multi-band colors into a finite set of galaxy phenotypes. Each phenotype
corresponds to a population of galaxies with similar observed colors and, in
general, similar redshift distributions. SOMPZ then calibrates the ensemble
redshift distribution of each phenotype using a limited redshift calibration
sample. In the idealized limiting case where each phenotype has an
infinitesimally narrow redshift distribution, a single calibration galaxy would
be sufficient to determine the redshift distribution of that phenotype. In
practice, SOMPZ performance is therefore controlled by how well the SOM
phenotypes localize galaxies in redshift. This localization depends on the noise
in the photometric measurements, the number of available passbands, and the
wavelength coverage.

For Roman-HLIS, these considerations naturally lead to a two-stage SOMPZ
framework. The medium-tier sample provides the wide-field galaxy sample whose
redshift distribution, $n(z)$, we want to estimate, but it contains only
three-band photometry. The deep tier provides seven-band photometry that is
approximately $1\,\mathrm{mag}$ deeper, and therefore provides more information about galaxy redshift given the measured colors. Finally,
only a much smaller subset of galaxies will have spectroscopic redshifts. The
goal of \texttt{Roman SOMPZ} is therefore to combine these three complementary
data sets: a wide sample with limited photometric information, a deeper
multi-band sample with more complete color coverage, and a redshift calibration
sample. 

The pipeline therefore uses two SOMs. The first is a \emph{wide SOM}, trained on the colors
of the wide-field galaxy sample. This SOM describes the galaxy population for
which we want to estimate $n(z)$, but by itself it has limited ability to break
color--redshift degeneracies because the wide data have fewer bands and noisier
photometry. The second is a \emph{deep SOM}, trained on deeper multi-band photometry with broader wavelength coverage. The deep SOM provides a finer
description of galaxy phenotypes and is better able to separate galaxies that
have similar wide-field colors but different redshifts. Once the SOMs are trained, galaxies are assigned to the cells whose representative colors most closely match their measured colors. We denote wide SOM cells by \(\hat{c}\) and deep SOM cells by \(c\). Galaxies in the wide-field sample are assigned to the wide SOM, giving the probability \(p(\hat{c})\) that
a galaxy occupies wide cell \(\hat{c}\). The deep sample and the redshift
calibration sample are assigned to the deep SOM. The redshift calibration sample
then provides an estimate of the redshift distribution associated with each deep
cell, \(p(z|c)\).

The key step in SOMPZ is to connect the deep SOM to the wide SOM through the
transfer function, $p(c|\hat{c})$, which gives the probability that a galaxy
assigned to wide SOM cell $\hat{c}$ would be assigned to deep SOM cell $c$ if the deep multi-band photometry were available. This transfer function enables
redshift information calibrated in the deep photometric space to be propagated
to the wide-field galaxy sample. In a typical cosmological analysis, $p(c|\hat{c})$ is
estimated using synthetic source-injection (SSI) simulations, in which galaxies
from the deep tier are injected into images representative of the wide-field
sample, with the appropriate noise properties and observing conditions applied,
and then processed with the same detection and photometry pipeline used for the
wide data. This procedure determines how galaxies with deep-tier properties
would be observed in the wide data and assigned to the wide SOM. For analyses focused on Roman deep-tier observations, the construction can be simplified
because the deep and wide samples fully overlap on the sky: the transfer
function can be estimated by measuring the same galaxies under photometric
configurations corresponding to the deep and wide samples, and then using these
matched measurements to map galaxies between the two SOMs. In this study, we
make a further simplifying approximation because only one set of photometric
measurements is available. We therefore use the same photometry to assign
galaxies to both the deep and wide SOMs, effectively removing additional
photometric-transfer effects from the analysis.

The redshift distribution of the wide sample can then be written as
\begin{equation}
p(z_{\rm wide}) =
\sum_{\hat{c}} \sum_c p(z|c,\hat{c})\,p(c|\hat{c})\,p(\hat{c}) .
\end{equation}
In practice, the redshift calibration sample is finite, and not every pair of
deep and wide SOM cells contains enough calibration galaxies to directly
estimate \(p(z|c,\hat{c})\). We therefore use the approximation that, at a fixed
deep SOM cell, the redshift distribution is only weakly dependent on the noisy
wide-field photometry. This assumption motivates the bin-conditional
approximation used in the SOMPZ estimator.

Tomographic redshift bins are defined using the wide SOM. Each wide SOM cell is
assigned to a tomographic bin based on the mode of the redshifts of the redshift
calibration galaxies associated with that wide SOM cell. The redshift
distribution of tomographic bin \(b\) is then obtained by summing over the wide
cells assigned to that bin and over the corresponding deep cells,
\begin{equation}
p(z|b) \approx
\sum_{\hat{c} \in b} \sum_c
p(z|c,b)\,p(c|\hat{c})\,p(\hat{c}|b).
\end{equation}

The final SOMPZ products are not only a single best-fit estimate of \(n(z)\), but an ensemble of possible redshift distributions that reflects the uncertainties. These include imperfections in the redshift calibration sample, photometric zero-point uncertainty in the deep fields, sample variance, shot noise from the finite number of calibration galaxies, and blending-related biases. Propagating these uncertainties produces a set of \(n(z)\) realizations for each tomographic bin, which can then be carried into the cosmological analysis. In this analysis, we simply include the sample variance equivalent to $20~\deg^2$ area to simplify the process.

\section{Photometry performance}
Figure \ref{fig:color_nndistance} shows the differences between measured colors and true colors in Roman and Rubin bands as a function of distance to nearest neighbors and figure~\ref{fig:color_nndistance_iH} shows the color differences for i-H158 measured in different methods. 

\begin{figure*}
    \centering
\includegraphics[width=0.81\linewidth]{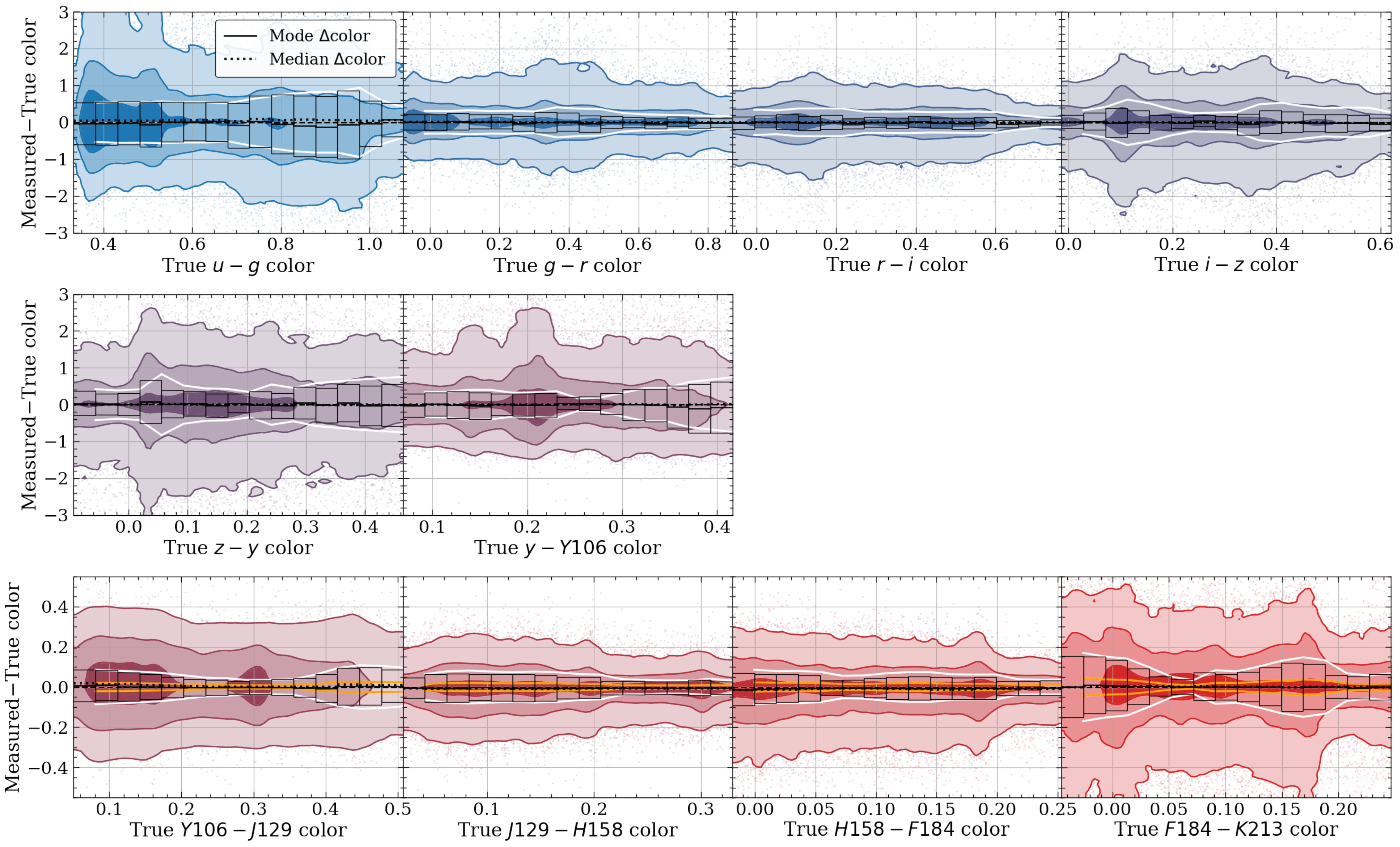}
    \caption{
Same as figure~\ref{fig:color_short}, but with all Rubin and Roman bands.
    }
    \label{fig:color}
\end{figure*}
\begin{figure*}
    \centering
\includegraphics[width=0.81\linewidth]{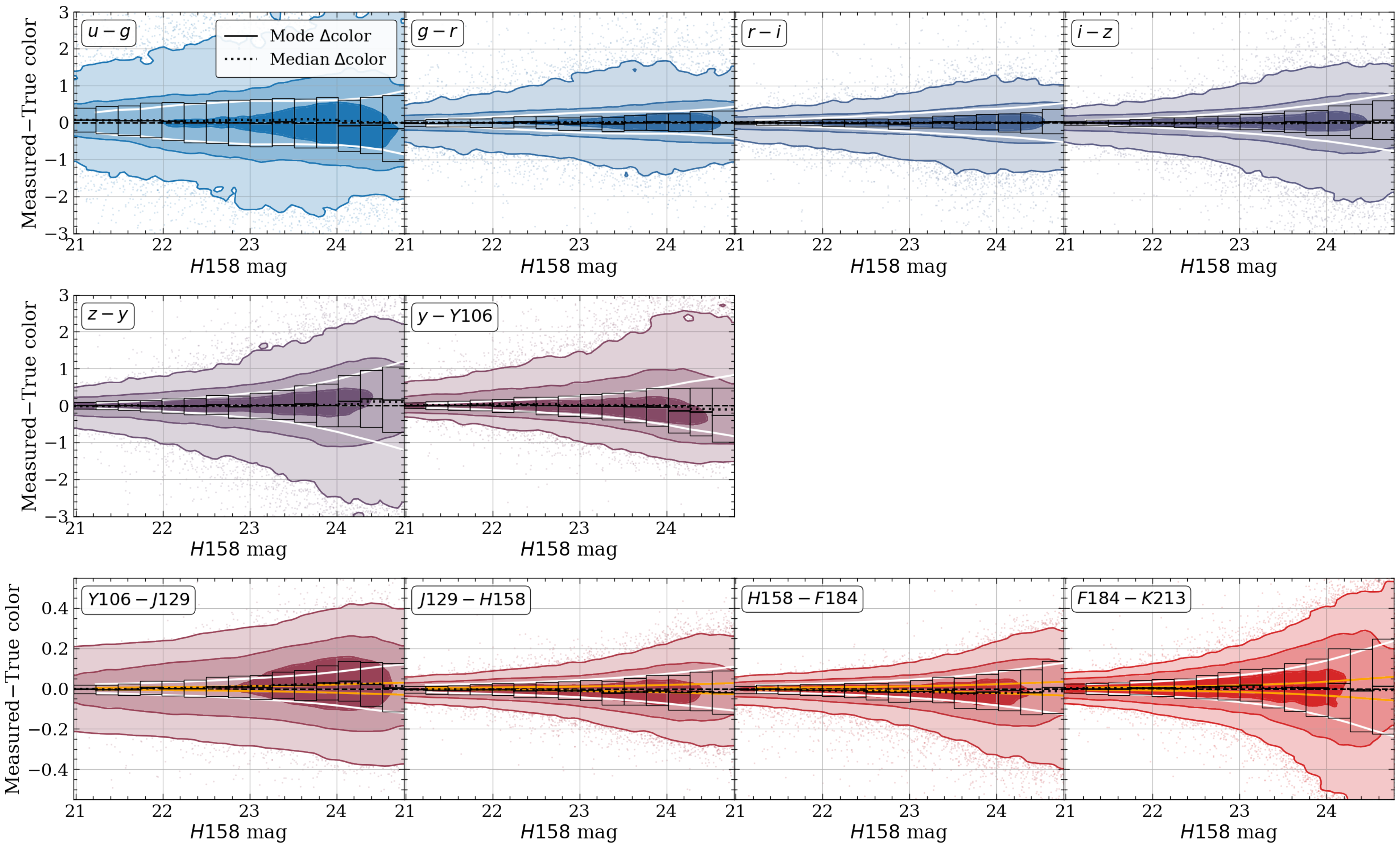}
    \caption{
Same as figure~\ref{fig:colormagnitude_short}, but with all Rubin and Roman bands.
    }
    \label{fig:colormagnitude}
\end{figure*}

\begin{figure*}
    \centering
    \includegraphics[width=1.0\linewidth]{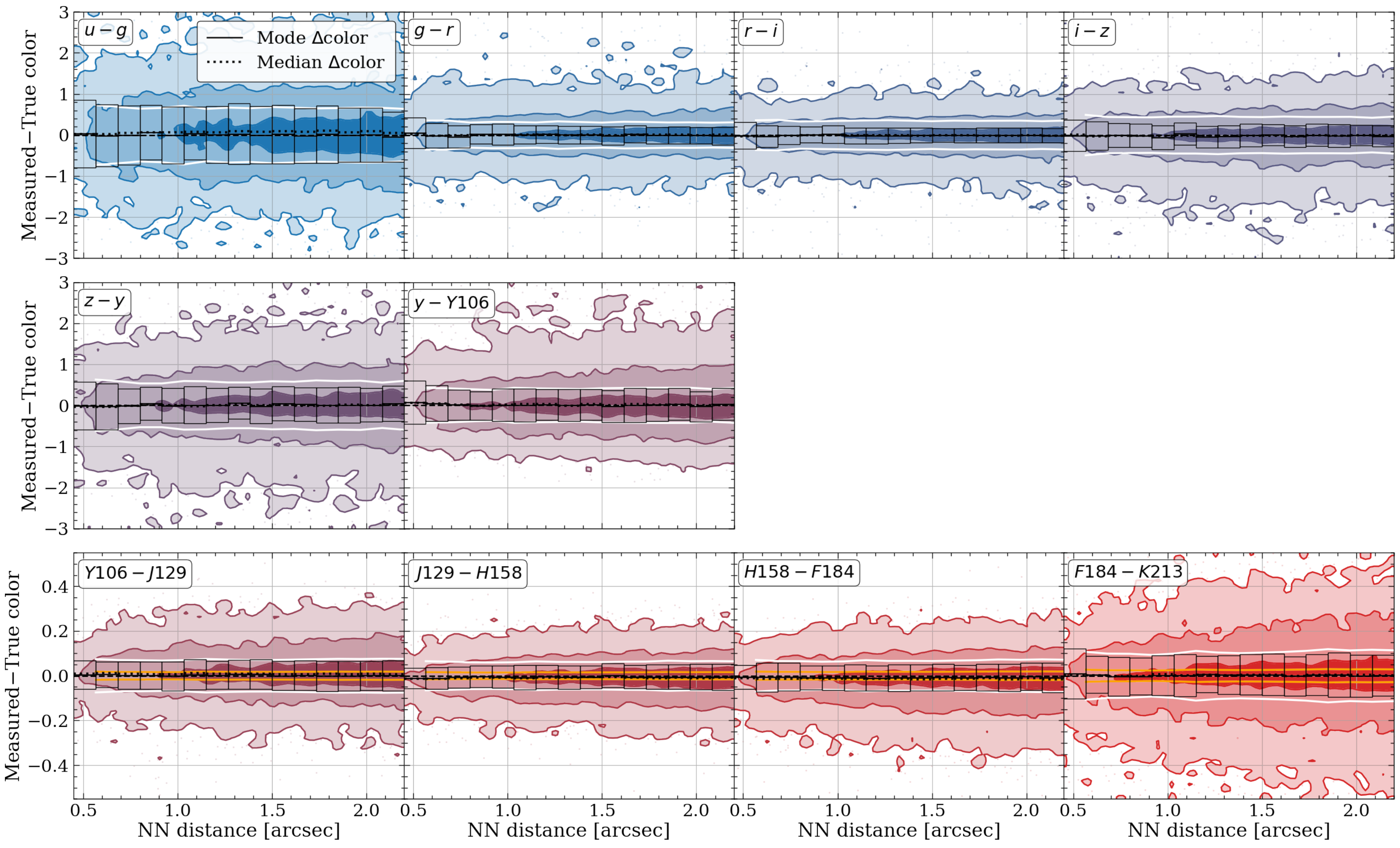}
    \caption{
Same as figure~\ref{fig:color}, but the x-axis changes to the distance to the nearest neighbors. The summary statistics are computed in 15 equal-sized bins from the $0.45''$ to the $2.2''$. 
    }
    \label{fig:color_nndistance}
\end{figure*}

\begin{figure*}
    \centering
    \includegraphics[width=1.0\linewidth]{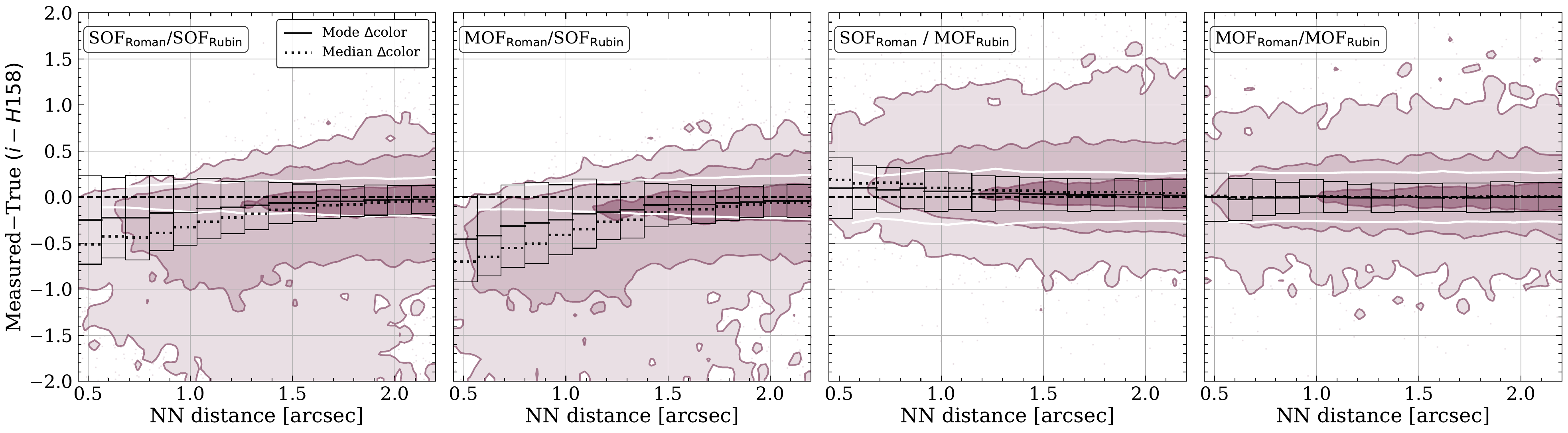}
    \caption{
Same as figure~\ref{fig:color_nndistance_different}, but for i-H158 color.
    }
    \label{fig:color_nndistance_iH}
\end{figure*}

\section{Spurious detections}
\begin{figure*}
    \centering
    \includegraphics[width=1.0\linewidth]{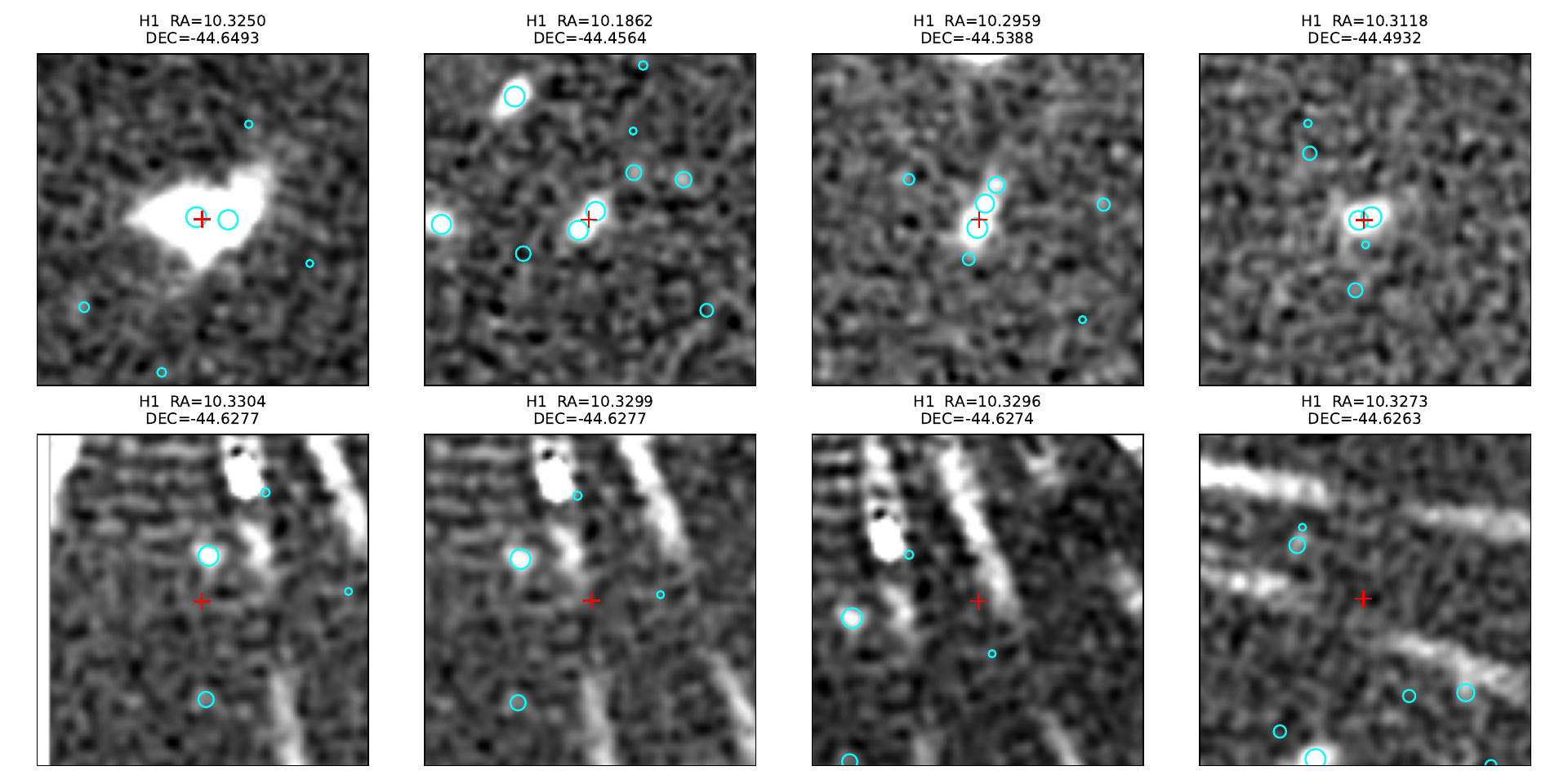}
    \caption{
Example $5''\times5''$ cutouts of detections that are not matched to the truth catalog in section~\ref{sec:absolutephotometyandcolor}. The grayscale image shows the Roman H-band image. Red crosses mark detected sources, while cyan crosses mark true galaxies. The top row shows examples from the subset of unmatched detections that have a truth galaxy within $0.147''$ but fail the magnitude-matching criterion, with a five-band magnitude distance greater than $1.0$; these cases account for about $73\%$ of the unmatched sample. The bottom row shows examples from the remaining $27\%$, which have no truth galaxy within $0.147''$ and are primarily associated with spurious detections caused by imperfect stellar masks.
    }
    \label{fig:bad}
\end{figure*}

As a diagnostic of detection failures and catastrophic photometric failures, we examine \slimfarmer{} detections that do not have a reliable match in the truth catalog in section~\ref{sec:absolutephotometyandcolor}. We define a detection as matched if there is a truth galaxy within $0.147''$ whose five-band magnitude distance, computed using the Y106, J129, H158, F184, and K213 bands, is smaller than $1.0$. With this criterion, $\simeq 4\%$ of the \slimfarmer{} detections remain unmatched. Figure~\ref{fig:bad} shows representative cutouts of these unmatched detections. Visual inspection indicates that most of these cases are associated either with severe blending, where a detected source cannot be cleanly associated with a single truth galaxy, or with spurious detections caused by imperfect stellar masks.

\section{Detection Configuration}

\label{app:detection}
The configuration of SEP used in this paper is detailed in table~\ref{tab:sep_detection_config}.

\section{Tabulated comparison}
Table~\ref{tab:photometry_accuracy} shows tabulated version of figure~\ref{fig:mag} and table~\ref{tab:colornndistance} shows tabulated version of figure~\ref{fig:color_nndistance}. 

\begin{table}
    \centering
    \caption{Configuration parameters used for source detection with \texttt{SEP}. The \texttt{gauss\_3.0\_7x7.conv} denotes a $7\times7$ matrix with FWHM of 3 pixels.}
    \label{tab:sep_detection_config}
    \begin{tabular}{ll}
        \toprule
        Parameter & Value \\
        \midrule
        \texttt{thresh}                & \texttt{3.0} \\
        \texttt{minarea}               & \texttt{5} \\
        \texttt{back\_bw}              & \texttt{32} \\
        \texttt{back\_bh}              & \texttt{32} \\
        \texttt{back\_fw}              & \texttt{2} \\
        \texttt{back\_fh}              & \texttt{2} \\
        \texttt{filter\_kernel}         & \texttt{gauss\_3.0\_7x7.conv} \\
        \texttt{filter\_type}           & \texttt{matched} \\
        \texttt{deblend\_nthresh}       & \texttt{$2^5$} \\
        \texttt{deblend\_cont}          & \texttt{0.001} \\
        \texttt{clean}                  & \texttt{True} \\
        \texttt{clean\_param}           & \texttt{1.0} \\
        \texttt{pixstack\_size}         & \texttt{20,000,000} \\
        \texttt{use\_detection\_weight} & \texttt{True} \\
        \bottomrule
    \end{tabular}
\end{table}

\begin{table*}
    \centering
    \begin{tabular}{|c|c|c|c|c|c|}
    \hline
    Magnitude & \multicolumn{5}{c|}{(mmag)}\\
    \hline
    & $Y106$ & $J129$ & $H158$ & $F184$ & $K213$ \\
    \hline
    17.00 -- 17.25 & 26 (29) & 12 (30) & 12 (28) & 17 (24) & 12 (27) \\
    17.25 -- 17.50 & 8 (23) & 8 (24) & 18 (25) & 18 (28) & 14 (26) \\
    17.50 -- 17.75 & 9 (32) & 13 (26) & 11 (26) & 21 (27) & 11 (27) \\
    17.75 -- 18.00 & 7 (28) & 5 (26) & 10 (23) & 15 (21) & 8 (21) \\
    18.00 -- 18.25 & 8 (24) & 8 (20) & 8 (23) & 14 (29) & 4 (23) \\
    18.25 -- 18.50 & 8 (28) & 6 (26) & 5 (21) & 13 (23) & 11 (25) \\
    18.50 -- 18.75 & 0.1 (24) & 5 (23) & 9 (25) & 13 (24) & 8 (27) \\
    18.75 -- 19.00 & 2 (22) & 3 (24) & 3 (29) & 8 (27) & 3 (26) \\
    19.00 -- 19.25 & 1 (27) & 0.2 (28) & -0.7 (26) & 6 (27) & 2 (28) \\
    19.25 -- 19.50 & -1 (26) & -3 (27) & -0.5 (25) & 4 (27) & -2 (29) \\
    19.50 -- 19.75 & -3 (26) & -3 (28) & -3 (29) & 2 (30) & -1 (31) \\
    19.75 -- 20.00 & -5 (29) & -4 (29) & -5 (31) & -2 (30) & -4 (34) \\
    20.00 -- 20.25 & -8 (30) & -7 (31) & -7 (31) & -4 (34) & -6 (35) \\
    20.25 -- 20.50 & -9 (32) & -9 (33) & -8 (34) & -6 (35) & -8 (39) \\
    20.50 -- 20.75 & -11 (31) & -11 (35) & -10 (36) & -9 (37) & -13 (43) \\
    20.75 -- 21.00 & -15 (39) & -12 (38) & -14 (38) & -11 (40) & -13 (44) \\
    21.00 -- 21.25 & -17 (38) & -16 (41) & -14 (40) & -12 (46) & -15 (50) \\
    21.25 -- 21.50 & -19 (42) & -18 (42) & -17 (45) & -15 (50) & -19 (56) \\
    21.50 -- 21.75 & -20 (45) & -22 (46) & -21 (51) & -19 (54) & -23 (62) \\
    21.75 -- 22.00 & -22 (48) & -26 (52) & -25 (54) & -20 (57) & -25 (67) \\
    22.00 -- 22.25 & -26 (51) & -29 (57) & -28 (60) & -23 (68) & -31 (79) \\
    22.25 -- 22.50 & -26 (57) & -32 (59) & -33 (67) & -27 (75) & -38 (89) \\
    22.50 -- 22.75 & -27 (62) & -37 (68) & -39 (74) & -31 (85) & -43 (97) \\
    22.75 -- 23.00 & -30 (69) & -42 (73) & -43 (82) & -33 (89) & -44 (106) \\
    23.00 -- 23.25 & -32 (78) & -52 (87) & -47 (89) & -34 (98) & -45 (120) \\
    23.25 -- 23.50 & -40 (88) & -60 (93) & -49 (95) & -35 (105) & -47 (134) \\
    23.50 -- 23.75 & -48 (99) & -71 (101) & -60 (105) & -41 (114) & -44 (149) \\
    23.75 -- 24.00 & -53 (106) & -79 (106) & -65 (110) & -45 (123) & -46 (166) \\
    24.00 -- 24.25 & -52 (116) & -83 (119) & -66 (123) & -51 (138) & -51 (191) \\
    24.25 -- 24.50 & -57 (130) & -90 (132) & -69 (135) & -50 (152) & -36 (213) \\
    24.50 -- 24.75 & -60 (141) & -80 (145) & -64 (145) & -41 (164) & -40 (233) \\
    24.75 -- 25.00 & -59 (155) & -69 (154) & -46 (149) & -29 (179) & -42 (260) \\
    \hline
    \end{tabular}
    \vspace{10pt}
    \caption{The median difference between the true and measured magnitudes as shown in figure \ref{fig:mag}, in units of millimags. The numbers in the parentheses show the standard deviation of the differences, shown as black bars in the aforementioned figure.}
    \label{tab:photometry_accuracy}
\end{table*}

\begin{table*}
    \centering
    \scriptsize
    \begin{tabular}{|c|c|c|c|c|c|c|c|c|c|c|}
    \hline
    NN dist [arcsec] & \multicolumn{10}{c|}{Color residual (mmag)}\\
    \hline
    & $u-g$ & $g-r$ & $r-i$ & $i-z$ & $z-y$ & $y-Y106$ & $Y106-J129$ & $J129-H158$ & $H158-F184$ & $F184-K213$ \\
    \hline
    0.450 -- 0.567 & 28 (823) & 27 (374) & -24 (326) & 11 (377) & -33 (581) & 64 (525) & 9 (64) & -13 (46) & -12 (54) & 0.8 (112) \\
    0.567 -- 0.683 & 23 (751) & -13 (292) & -15 (267) & -23 (352) & -44 (560) & 40 (435) & 11 (58) & -13 (48) & -13 (64) & 1 (82) \\
    0.683 -- 0.800 & 48 (720) & 0.0 (268) & -12 (222) & -19 (293) & -14 (425) & 14 (367) & 10 (61) & -11 (47) & -13 (63) & -2 (87) \\
    0.800 -- 0.917 & 73 (666) & -2 (233) & 0.7 (192) & -10 (287) & 20 (401) & -14 (343) & 8 (60) & -9 (48) & -11 (53) & -2 (83) \\
    0.917 -- 1.033 & 68 (694) & -2 (250) & 3 (218) & -18 (285) & 5 (422) & 13 (393) & 6 (64) & -7 (49) & -15 (55) & 6 (89) \\
    1.033 -- 1.150 & 74 (646) & 5 (222) & -8 (189) & -14 (267) & -26 (454) & 36 (405) & 13 (68) & -11 (55) & -13 (58) & 0.8 (94) \\
    1.150 -- 1.267 & 73 (699) & 6 (219) & -5 (189) & -19 (259) & -3 (418) & 29 (375) & 9 (62) & -10 (52) & -14 (56) & 5 (81) \\
    1.267 -- 1.383 & 90 (740) & 3 (219) & -0.7 (184) & -17 (268) & 27 (389) & -14 (390) & 8 (65) & -8 (53) & -14 (60) & 2 (86) \\
    1.383 -- 1.500 & 75 (695) & 8 (215) & -10 (181) & -14 (274) & -0.2 (423) & 27 (369) & 10 (67) & -6 (52) & -15 (59) & 5 (90) \\
    1.500 -- 1.617 & 92 (711) & -2 (222) & -7 (169) & -12 (258) & 14 (420) & 4 (383) & 7 (65) & -6 (54) & -12 (60) & 4 (93) \\
    1.617 -- 1.733 & 83 (661) & 10 (207) & 1 (174) & -24 (265) & -6 (406) & 30 (415) & 8 (66) & -8 (53) & -13 (56) & -1 (87) \\
    1.733 -- 1.850 & 105 (702) & 8 (200) & -13 (177) & -3 (267) & -14 (427) & 20 (400) & 10 (67) & -12 (55) & -12 (60) & 3 (91) \\
    1.850 -- 1.967 & 66 (662) & 15 (211) & -12 (183) & -13 (274) & 11 (437) & 20 (394) & 8 (67) & -9 (55) & -15 (61) & 4 (98) \\
    1.967 -- 2.083 & 98 (663) & -0.5 (196) & -8 (179) & -3 (246) & -3 (407) & -3 (365) & 10 (65) & -11 (57) & -11 (61) & 4 (94) \\
    2.083 -- 2.200 & 53 (615) & 6 (218) & -10 (172) & -7 (256) & 16 (428) & 8 (390) & 8 (68) & -10 (58) & -17 (66) & 1 (92) \\
    \hline
    \end{tabular}
    \caption{Median color residual (measured$-$true) vs. nearest-neighbour distance for all Rubin and Roman bands as shown in figure \ref{fig:color_nndistance}. Values are in mmag; parentheses show the standard deviation of the differences, which are shown as black bars in the aforementioned figure.}
    \label{tab:colornndistance}
\end{table*}

\section{Origin of the photometric residuals at faint magnitudes}
\label{app:magnitude_residual}
One possible explanation for the negative magnitude residuals at the faint end is selection bias near the detection threshold. At a fixed true magnitude, galaxies with positive flux fluctuations are more likely to be detected, while those with negative flux fluctuations may fall below the detection threshold and be missed. This truncation would preferentially retain galaxies whose measured fluxes are biased high, shifting the magnitude residuals toward more negative values at faint magnitudes.

\begin{figure*}
    \centering
    \includegraphics[width=0.5\linewidth]{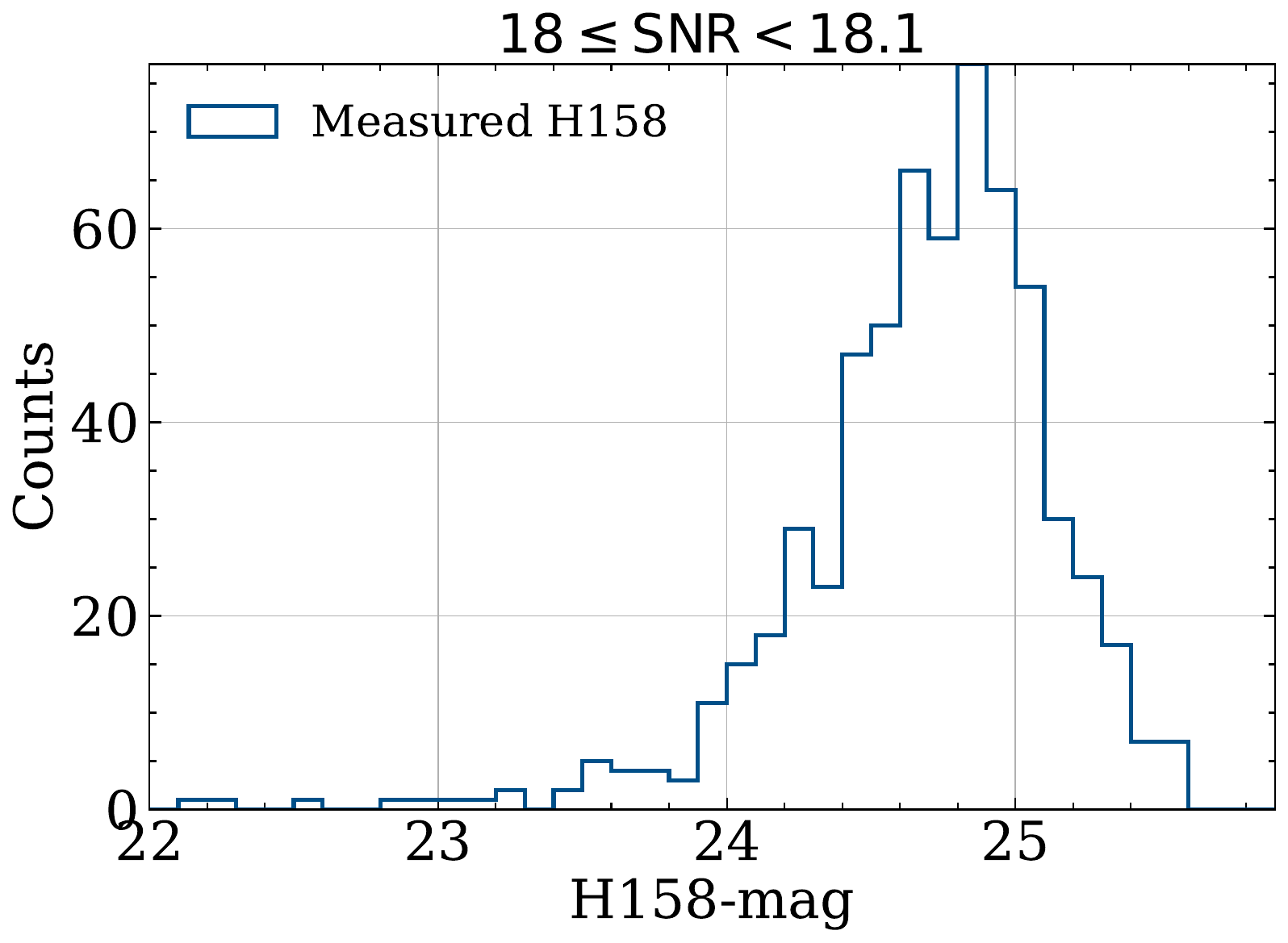}
    \caption{The histogram of H158 magnitude distribution for galaxies around the detection threshold. 
    }
    \label{fig:detectionSNR}
\end{figure*}

\begin{figure*}
    \centering
    \includegraphics[width=1\linewidth]{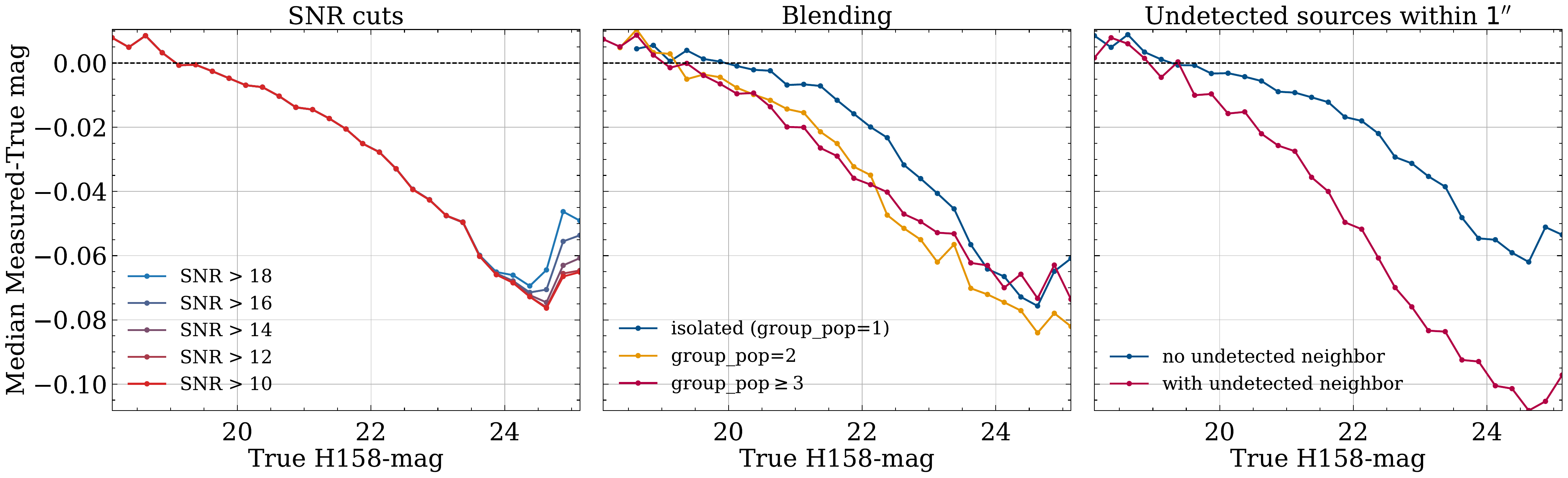}
    \caption{Distribution of the difference between the measured galaxy H158 magnitude and the true galaxy H158 magnitude as a function of true H158 magnitude for the galaxy sample.  Left: median residual for samples selected with mean combined-flux signal-to-noise thresholds ranging from SNR $> 10$ to SNR $> 18$.  Middle: median residual split by the number of sources fit simultaneously in the
modeling. Right: median residual for galaxies with and without an undetected
  truth source within $1''$.}
    \label{fig:bias_summary}
\end{figure*}

However, our tests suggest that this effect is unlikely to be the primary cause. In figure~\ref{fig:detectionSNR}, we show the magnitude distribution of galaxies near the detection threshold and find that it lies mostly beyond the magnitude range where the residuals begin to deviate. In the first panel of figure~\ref{fig:bias_summary}, we further show that the bias remains nearly unchanged when the signal-to-noise cut is varied from 10 to 18. These results suggest that the faint-end residuals are not primarily driven by selection effects near the detection threshold.

Instead, the residuals appear to be more closely associated with flux contamination from neighboring sources. In the second panel of figure~\ref{fig:bias_summary}, we split the sample according to whether each galaxy has nearby detected neighbors and find that galaxies without detected neighbors have smaller residuals. In the third panel of figure~\ref{fig:bias_summary}, we perform a similar split using true galaxy neighbors within $1''$ and again find smaller residuals for galaxies without nearby truth objects. This difference remains qualitatively unchanged when the neighbor aperture is increased to $2''$ or $3''$. These findings suggest that the residuals are likely caused by flux contamination from nearby detected and undetected sources, including unmodeled light from galaxies below the detection threshold and residual deblending errors in multi-galaxy groups.

\end{appendix}

\end{document}